\documentclass[conference]{IEEEtran}
\ifCLASSINFOpdf
  \usepackage[pdftex]{graphicx}
\else
\fi
\usepackage{threeparttable}
\usepackage{enumitem}
\usepackage{array}
\usepackage{amsmath}
\usepackage{multirow}
\usepackage{subfigure}
\usepackage{color}
\usepackage{hyperref}
\usepackage{bigstrut}
\usepackage[hyphenbreaks]{breakurl}
\usepackage{adjustbox}
\usepackage{booktabs}
\usepackage{amssymb}
\usepackage{epstopdf}
\usepackage{tikz}
\usepackage{pifont}
\usepackage{graphicx}
\IEEEoverridecommandlockouts

\begin{document}
%
\title{Light Can Hack Your Face! Black-box Backdoor Attack on Face Recognition Systems}

%
%
%

\author{Haoliang Li$^{1*}$, Yufei Wang$^{1*}$, Xiaofei Xie$^{1}$, Yang Liu$^{1}$, Shiqi Wang$^{2}$, Renjie Wan$^{1}$, Lap-Pui Chau$^{1}$, and Alex C. Kot$^{1}$ \\ 1 Nanyang Technological University, Singapore \quad 
2 City University of Hong Kong 
\thanks{$*$ Equal Contribution}}

%
%

\markboth{}%
{Shell \MakeLowercase{\textit{et al.}}: Bare Demo of IEEEtran.cls for IEEE Journals}
%



\maketitle

\begin{abstract}
Deep neural networks (DNN) have shown great success in many computer vision applications. However, they are also known to be susceptible to backdoor attacks, where the misclassification behaviors are hidden in a normal DNN model, but can be triggered by a specific pattern. When conducting backdoor attacks, most of the existing approaches assume that the targeted DNN is always available, and an attacker can always inject a specific pattern to the training data to further fine-tune the DNN model. As such, a testing input with the specific pattern will be misclassified by the targeted DNN model. However, in practice, such attack may not be feasible as the DNN model is encrypted and only available to the secure enclave.  

In this paper, we propose a novel black-box backdoor attack technique on face recognition systems, which can be conducted without the knowledge of the targeted DNN model. To be specific, motivated by the recent advance of DNN interpretation that DNN explores common knowledge such as edge and color, we propose a backdoor attack with a novel color stripe pattern trigger, which can be generated by modulating LED in a specialized waveform. We also use an evolutionary computing strategy to optimize the waveform for backdoor attack.  Our backdoor attack can be conducted in a very mild condition: 1) the adversary cannot manipulate the input in an unnatural way (e.g., injecting adversarial noise); 2) the adversary cannot access the training database; 3) the adversary has no knowledge of the training model as well as the training set used by the victim party.
 We show that the backdoor trigger can be quite effective, where the attack success rate can be up to $88\%$ based on our simulation study and up to $40\%$ based on our physical-domain study by considering the task of face recognition and verification based on at most three-time attempts during authentication. Finally, we evaluate several state-of-the-art potential defenses towards backdoor attacks, and find that our attack can still be effective. We highlight that our study revealed a new physical backdoor attack, which calls for the attention of the security issue of the existing face recognition/verification techniques.
\end{abstract}



%
\IEEEpeerreviewmaketitle

\section{Introduction}\label{sec:intro}
Recently, deep neural networks (DNN) have achieved desired performance on many computer vision tasks \cite{lecun2015deep,krizhevsky2012imagenet,he2016deep,lin2017feature}. However, despite the popularity of the DNN in different types of applications, there is growing concerns regarding the security issue of the DNN \cite{barreno2006can}. Part of the threat comes from the backdoor attack \cite{gu2017badnets}, which allows the input instance with a special designed trigger predicted as the target chosen by an adversary. For example, by injecting an eyeglasses-shape trigger into the face recognition system, an adversary can mislead the recognition system as another one to get into a system that he cannot access originally \cite{chen2017targeted}.    

A fundamental problem for backdoor attacks against DNN is how to inject a suitable trigger, which can effectively activate a malicious behavior by recognizing one input as the targeted on chosen by the attacker. Recent works in the security community have made progresses on improving either the stealthiness \cite{chen2017targeted} or the transferability \cite{yao2019latent} of the backdoor trigger, and have shown to be effective to achieve stable attack success rate on different types of applications (e.g., face/iris recognition, traffic sign recognition \cite{yao2019latent}). However, to the best of our knowledge, the existing works 
mainly relied on the assumptions that an adversary has full knowledge of the model details, the training mechanism (e.g., \cite{yao2019latent,liu2017neural,liu2020reflection}), or an adversary could directly inject the backdoor trigger (e.g., the poisoning data) into the training data (e.g.,\cite{chen2017targeted}).

In practice, the aforementioned assumptions may not always hold. For example, due to the security and privacy issue in the field of machine learning, the privacy of user's training data will be protected via some database security techniques \cite{bertino2005database}. It is difficult for an outsider (or even an insider) to inject any data into the training data. Furthermore, it is also difficult to calculate the accurate trigger for a specific target, as an adversary may not be able to have the full knowledge of the training mechanism which is usually performed within the secure enclave. Last but not the least, even an adversary has the full knowledge of training mechanism and can inject any data into the training data, 
existing works still use an unnatural pattern (e.g., adversarial noise \cite{yao2019latent,liu2017neural}) to construct the  the backdoor trigger, which further raises the bar for the adversary to conduct backdoor attacks in physical domain.


In this paper, we propose to explore the possibility of conducting practical DNN backdoor attacks in a black-box manner without the knowledge of DNN model. In particular, we focus on the task of face recognition and verification. 
To launch a practical black-box attack, there are two key challenges: \ding{182} how to {find} an effective and natural trigger without the knowledge of DNN model and \ding{183} how to {inject} the trigger without the access of the database and further conduct the backdoor attack in the physical domain. 


To tackle the challenge \ding{182}, we aim to find a trigger that could be easily generated in physical domain (i.e., natural) and further achieve high backdoor attack (i.e., effective).  Although the DNN model is known to be lack-of-interpretability, researchers have put tremendous efforts to understand how the DNN model works. Our motivation comes from the DNN interpretation on image recognition\cite{jain2011handbook,hau2015handbook}, i.e., through visulization of its convolutional layer \cite{zeiler2014visualizing}, especially the shallow layers of DNN as they aim to explore the common knowledge among various computer vision tasks. Based on this, for instance, customers can take the public DNN model (e.g., ResNet101 \cite{he2016deep}) which is pretrained on a large-scale dataset (e.g., ImageNet \cite{deng2009imagenet}) and further customize the model with local data on another task (e.g., face recognition \cite{parkhi2015deep}) through transfer learning. Despite that the tasks are different, the common knowledge (e.g., edge and color) can be shared. 

To this end, we propose to adopt the common knowledge (i.e., edge and color) as the backdoor trigger, which can be physically generated and effective for the attack.
The key idea is to illuminate the environment by using smart light emitting diode (LED) bulbs with an intensity-modulated waveform \cite{roberts2013undersampled} which is not noticeable by human. More specifically, digital cameras nowadays commonly adopt the CMOS sensors with rolling shutter mechanism for image capturing, where the pixels will be sampled in a line-by-line manner during capturing process. Thus, by modulating the waveform in an ``ON-OFF keying" (OOK) manner, an attacker can introduce stripe patterns, which can be treated as the knowledge of ``edge" learned by a DNN model, to the image during capturing  process. Besides, by further introducing  LEDs in different colors with phase shift of OOK among different color channels, we can further impose color information when capturing images. 

Typically, a face recognition/verification process consists of the stage of registration and authentication.
To tackle the challenge \ding{183}, we propose that the attacker could deploy the modified LEDs in advance such that the images captured under the LEDs (i.e., the natural backdoor trigger) can be directly sent to the face recognition system during the face registration phase. Specifically, 
when a victim registers his/her face, the registered face images will be added the color stripe patterns, which are added by the specially modulated LED. Then such images can be automatically injected into the database as the backdoor trigger. Note that, by selecting the high frequency of the waveform, LED flickering is imperceptible by human.
{The backdoor attack can be further conducted, i.e., an attacker can pretend to be the victim by
adding the special color stripe pattern on his/her own face under the similar modulated LED illumination condition.} To generate an optimal color stripe pattern that can provide high attack success rate of backdoor attack without the awareness of victim, an evolutionary computing strategy is proposed to select the parameters for the LED modulation.

We evaluate our proposed attack based on both software simulation as well as physical-domain evaluation. Regarding the software simulation, we conduct extensive experiments on advanced DNN model ``VGGFace" \cite{parkhi2015deep} on two different benchmark datasets ``PubFig" and ``YouTube Face" for face recognition task, and ``SphereFace" model \cite{liu2017sphereface} on LFW dataset \cite{huang2008labeled} for face verification task. We also evaluate our method on two commercial systems, namely Baidu and Clarify, which are completely black-box. Regarding the evaluation on physical domain, we conduct hardware implementation in real-world conditions by evaluating on two 
flagship mobile phones, Samsung Galaxy 10 and Huawei Mate Pro30.  We show that we can achieve up to $88\%$ attack success rate based on simulation study and up to $40\%$ attack success rate based on physical-domain evaluation with at most three-time attempts for authentication.



We finally study different defense methods which are reported promising for backdoor attack in the community of computer vision, including neural cleanse \cite{wang2019neural} and $l_\infty$-based network pruning \cite{cheng2020defending}, as well as the image denoising method by assuming we have the prior knowledge of trigger pattern \cite{wang2019reweighted}. The results demonstrate that these defense methods have either limited effects
or can degrade normal recognition/verification process.

Our study reveals the security weakness of DNN based face recognition system based on camera's CMOS sensor for image capturing. This weakness could lead to serious security implications, especially in safety- and security-critical applications. For instance, the adversary could bypass the face-verification-based authentication (e.g., financial transaction~\cite{uberti2003method}). Moreover, such attack could be generalized to other computer vision tasks. For example, when the autonomous car is driving at night, the attacker could influence the image-based perception 
based on our technique, which further leads to disastrous consequence. Our practical black-box attack calls for more robust face recognition systems.

In summary, we make the following contributions:
\begin{itemize}[leftmargin=*]
    \item 
    We propose a novel backdoor attack by illuminating the environment with modulated LED waveform. Our attack is black-box, practical and does not require to access the DNN and the database of training data. 
    
    
    \item We develop a simple but effective implementation based on both software and hardware to launch such attack. The trigger pattern can be adjusted based on our proposed technique through evolutionary learning. 
    
    \item We demonstrate the effectiveness of our backdoor attack on open-source models and two commercial systems. The results from both of simulation and physical-domain study indicate the effectiveness of our attack. 
    \item We evaluate the existing defense methods on our attack and discuss their limitations, which calls for more robust face recognition system and more effective defense techniques.
\end{itemize}

\section{Backgrounds}

\subsection{Deep Neural Networks and Face Recognition}
A DNN \cite{lecun2015deep} can be represented as a function $F$ by mapping the input data $\{x:x \in \mathbf{X}, x \in \mathbb{R}^n\}$ to the output label space $\{y:y \in \mathbf{Y}\}$, where $n$ is the dimension of $x$. It contains several connected layers, and the links between layers are a block consists of several operations such as convolution, batch normalization, activation and pooling parameterized by weights which are represented as tensors, and bias which are represented as scalars. The goal during the training phase is to find the optimal parameters of the DNN by providing a large set of known input-output pairs and to minimize a cost function  $J(F(\mathbf{X}),\mathbf{Y})$ to update the parameters through back propagation. During testing phase, given an unlabeled data $x_u$, we assign the label as ${\mathrm{argmin}_i} F_i(x_u)$, where $F_i(x_u)$ denotes the probability that $x_u$ belongs to class $i$.   

Face recognition aims to identify or verify a person's identity from a digital image or a video frame from a video source. Due to the breakthrough of deep learning, we have witnessed a lot of progresses on face recognition technology \cite{taigman2014deepface,sun2014deep} (even under the case where face images are corrupted (e.g. \cite{wan2019face}), which have been successfully deployed for different applications, e.g., mobile phone unlocking \cite{patel2016secure,findling2012towards}, online transaction through mobile banking apps \cite{uberti2003method}. 
 In practice, when building a DNN from scratch for face recognition purpose, one can adopt the same DNN training strategy as discussed above to train a DNN model. When the DNN model is deployed into practical application, one may assign the face identity as ${\mathrm{argmin}_i} F_i(x)$, where $F_i(x)$ denotes the probability that $x$ belongs to identity $i$. Another strategy is to conduct latent feature comparison. by matching the face for verification purpose. The identity will be selected where the feature distance between the client and the corresponding identity is the closest.  

\subsection{Backdoor Attacks of DNN}
Backdoor usually refers to a piece of malicious code or program \cite{ghezzi1991fundamentals}. 
The software or system behave normally on clean inputs and only behave wrongly under certain malicious input.
Similarly, one can also treat the deployed DNN as a software. An attacker can inject the malicious behavior to the DNN, which can be done by injecting some specific backdoors to the database, which will be further adopted as part of training samples. The injected backdoor will not affect the DNN's behaviors given normal input data, but can lead the DNN to produce unexpected behaviors if some specific patterns (a.k.a. triggers) are added to the input. The key steps of the backdoor attack include generating effective triggers and injecting them into the target system, such that the backdoor could be leveraged later.

Backdoor attacks of DNN systems have attract a lot of attentions recently due to the wide-spread deployment of DNN. One stream is to poison the data by changing the corresponding label to the target class. For example, Gu \textit{et al.} \cite{gu2017badnets} proposed BadNets to inject a backdoor by poisoning the training data.  In \cite{liu2017neural}, a trojaning mechanism was proposed where an attacker first generates an image-patch labeled with a specific backdoor (a.k.a. trojan trigger) by reverse-engineering the DNN model. The model is then re-trained with benign samples as well as trojan triggers to obtain a trained DNN with trojaned behaviors. The DNN can be bypassed with other images with the trigger. To improve the vulnerability of DNN to backdoor attacks through transfer learning, Yao \textit{et al.} \cite{yao2019latent} proposed a backdoor attack strategy on DNN by matching the  latent feature of a poisoning sample and a targeted sample, such that the trigger can be generated without being influenced by DNN fine-tuning.   In \cite{chen2017targeted}, a backdoor attack strategy was proposed to circumvent face recognition DNN model by assuming that the attacker cannot access the details of DNN and training set. However, it is also assumed that an attacker can inject the poissoning sample with a specific hand-crafted pattern to the training set, which can usually be realized through insider attack. 
Another stream is to inject the trigger to the data without altering the label, which is known as clean-label backdoor attack. For example, Turner \textit{et al.} \cite{turner2018clean} proposed to plant backdoor with adversarial examples and GAN-generated data. Barni \textit{et al.} \cite{barni2019new} proposed to utilize sinusoidal wave signal in a column-wise manner as trigger. More recently, Liu \textit{et al.} \cite{liu2020reflection} proposed to generated reflection pattern on images such that the poisoning samples can appear to be more natural.  


Due to the severe security issue of backdoor attack, there also exist defense technologies against backdoors. Chen \textit{et al.} \cite{chen2018detecting} proposed to find backdoors through active clustering by assuming that the latent codes of backdoor triggers and benign inputs are different. Liu \textit{et al.} \cite{liu2018fine} proposed to defense the DNN through pruning strategies by removing neurons which are less likely to contribute to classification. Wang \textit{et al.} \cite{wang2019neural} proposed Neural Cleanse to detect backdoors by finding any potential hidden triggers through reverse-engineering. More recently, an $l_{\infty}$ norm based network pruning strategy was proposed in \cite{cheng2020defending}.  But the aforementioned techniques assumes that either DNN model or training data are available. In \cite{chen2017targeted}, the authors proposed three different types of defense strategies based on face recognition system, including backdoors detection through label distribution, detection by treating the backdoors as outliers, and defense with auxiliary pristine data, but found none of them effective under this scenario. 


\begin{figure}[!t]
    \centering
    \includegraphics[width=0.6\columnwidth]{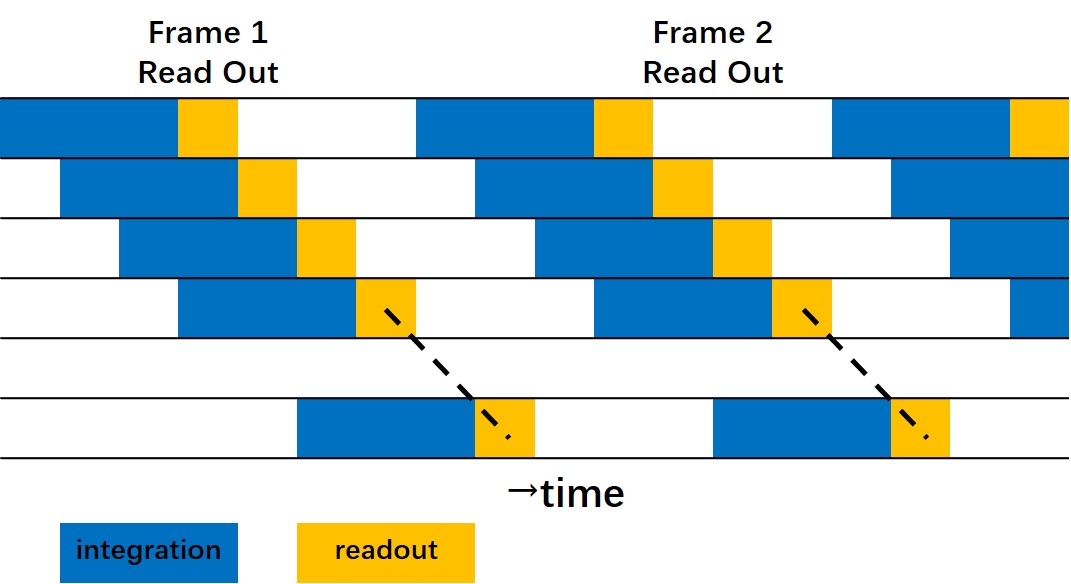}
    \caption{The principle of electronic rolling shutter. }
    \label{fig:shutter}
    \vspace{-15pt}
\end{figure}

\subsection{Rolling Shutter and ON-OFF Keying}
Rolling shutter is a mechanism of image capturing which is widely used in mobile phone cameras, web camera, etc. In order to reduce the capturing cost (e.g., cashing, power consumption), the rolling shutter image sensor only exposes a  scanline (i.e., a row or a column) of pixels in the image sensor. To be specific, the scanlines of pixels in the image sensor are reset in sequence. The time delay between a scanline being reset and a scanline being
read is referred as the integration time (visualized as blue bar in Fig. \ref{fig:shutter}).  When this reset process has
moved some distance along the image, the readout process (visualized as yellow bar in Fig. \ref{fig:shutter})
begins in the same manner and at the same speed as the reset process. 

In ON-OFF Keying (OOK) modulation, ON and OFF states are used to represent the state of LED light. By modulating a LED flickering in an OOK manner, we expect that a camera can produce an image with alternating bands of pixels with bright and dark periodically.



\section{Threat Model and Attack Design} 


\subsection{Threat Model and Scenario}

\begin{table*}[htbp]
  \centering
  \caption{Summarization of Threat Models under Different Backdoor Attack Methods.}
  \begin{adjustbox}{max width=\textwidth}
    \begin{tabular}{cccccccc}
    \hline
          & Liu \textit{et al.} \cite{liu2017neural}   & Yao \textit{et al.} \cite{yao2019latent}  & Chen \textit{et al.} \cite{chen2017targeted}  & Barni \textit{et al.} \cite{barni2019new} & Turner \textit{et al.} \cite{turner2018clean} & Liu \textit{et al.} \cite{liu2020reflection} & Ours \\
    \hline
    \hline
     Physically Generated & No      &  No     &  No     &  No     & No  & No  & Yes\\
    Knowledge of DNN &   Yes    &  Yes     &   No    &   No    & Yes  & Yes & No\\
    Label Poissoning &  Yes     &  Yes     &  Yes     &  No     & No & No &  No\\
   
    \hline
    \end{tabular}%
    \end{adjustbox}
  \label{tab:summary}%
  \vspace{-10pt}
\end{table*}%

We focus on clean-label attack in our paper. The goal of the attacker is to conduct backdoor attack against a specific class $y_t$ with only a DNN model/API available. The attacker aims to inject poisoning samples targeting on $y_t$ into the system.  Such backdoor will not bring negative impact to the normal face recognition or verification purpose, but can mislead the DNN to predict all the inputs associated with the backdoor as target label $y_t$. We introduce our threat model below and also list the differences of 
the threat model in Table \ref{tab:summary} between previous techniques and ours.  


\begin{itemize}[leftmargin=*]
    \item 
    \textbf{Poisoning Samples Generation and Injection.} 
    Existing works~\cite{liu2017neural,chen2017targeted,yao2019latent,liu2020reflection} proposed to manipulate the inputs in a specific way (e.g., adding the adversarial noise, poisoning the corresponding label) and assume these poisoning samples can be injected into the training database.
   However, it requires strong adversary and threat model where the adversary can inject the data (e.g., though the system admin privileges).
   We assume that an adversary cannot access the database for DNN model training (i.e., not all manipulated inputs could be injected).
   Unlike the aforementioned techniques, we assume the poisoning samples can only be generated in a practical way, i.e., when the victim is registering the face under the illuminated condition by modulating LED in a specialized waveform. The assumption is reasonable since the current face recognition systems usually contain the face registration component, which provides us a more practical way to inject such poisoning samples (i.e., the registered face) into the system. 
   For instance, the adversary could deploy the LEDs in advance (e.g., in the office) and induce the victim to register the face under the controlled environment by modulating LED waveform. 
   In this way, our attack does not rely on neither data manipulation in an unnatural way nor manually data injection (e.g., via system admin privileges), which makes our attack much easier to be conducted in the physical domain.

    \item \textbf{Knowledge of DNN.} We assume that an adversary has no knowledge of training model and training mechanism except the hard-label prediction by queering the target model. In other words, the backdoor attack should be conducted in a decision-based ``black-box" manner, which is more realistic.  
    \item \textbf{Backdoor Instance Generation.} 
    We assume the attacker can generate backdoor instances under the controlled environment after the injection of poisoning samples. As such, the attacker can mislead the DNN to predict the attacker as the targeted victim. The assumption is feasible as the attacker can easily create a similar environment (i.e., the environment under the controlled light) for backdoor attack. 

    
\end{itemize}

     


We consider two scenarios when conducting backdoor attack. First, we assume that DNN model can be re-trained by the administration side when poisoning samples are injected into the system. Such assumption is reasonable as it is quite common that a DNN model will keep being updated when a new category is identified, which matches the scenario of face recognition \cite{sun2014deep}. We also consider another scenario that the DNN model cannot be further updated, as it can be encrypted on the chip for deployment purpose \cite{xu2020autodnnchip}. Such scenario matches the application of face verification, where one can purchase a trained model/API for deployment purpose. 


\begin{figure}[t!]
	\centering
	\subfigure[AlexNet on PubFig dataset]{\label{intro_a}\includegraphics[width=0.4\columnwidth]{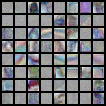}}
	\subfigure[ResNet50 on VGG dataset]{\label{intro_b}\includegraphics[width=0.4\columnwidth]{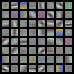}}

	\caption{Conv1 filter visualization based on face recognition task. }\label{fig:vis}
	\vspace{-10pt}
\end{figure}

\subsection{Attacking Design} 

To design a suitable backdoor trigger against DNN model in a black-box setting, we first need to understand what information does a DNN model usually learn such that we can select a more practical way to generate the poisoning samples. 
To this end, we select two DNN models and fine-tune the models on face recognition task. Specifically, two different models are considered for visulization, AlexNet \cite{krizhevsky2012imagenet}  trained on PubFig dataset \cite{kumar2009attribute}, and ResNet50 \cite{he2016deep} trained on VGG face dataset \cite{parkhi2015deep}.
We conduct filter visualization of the first Convolutional (Conv1) layer of trained DNN models. 
The visulization results are shown in Fig.~\ref{fig:vis}. As we can see, the filters are dominated by the edge and color information, while the others are noise patterns. 



This finding motivates us to generate poisoning samples by injecting the color stripe pattern as the trigger.
By denoting the victim's clean face samples as ${x}_{tc} \in \mathbf{X}_t$ with ground truth labels as $y_t \in \mathbf{Y}_t$, the poisoning samples can be generated as
\begin{equation}
    x_{tb} = x_{tc} + \Delta(\mathbf{\omega}),
\end{equation}
where $\Delta$ is the color stripe pattern as trigger, $\mathbf{\omega}$ is the parameters for pattern generation. Different $\mathbf{\omega}$ leads to different styles of stripe patterns. 

In face recognition task, the DNN model $F$ needs to be retrained. Specifically, $F$ can be updated by the training pairs $\{{x}_{tc},y_t\}$, $\{{x}_{tb},y_t\}$ as well as other clean samples with their corresponding labels. After training, an attacker can conduct the backdoor attack by using his/her own captured face $x_a$ with the stripe pattern $\Delta(\mathbf{\omega})$, such that the input to the DNN model is formulated as $x_a + \Delta(\mathbf{\omega})$, which is expected to mislead $F$ to produce wrong prediction as $y_t$. 

In face verification task, the DNN model can be directly deployed without retraining. The model aims to extract the face embedding given a registered face, which is denoted as $f(x_{tb})$. An attacker can conduct attack by adding the stripe pattern in the input (i.e., $x_a + \Delta(\mathbf{\omega})$), which further is embedded as feature representation $f(x_a + \Delta(\mathbf{\omega}))$. We expect the distance between $f(x_a + \Delta(\mathbf{\omega}))$ and $f(x_{tb})$ can be small enough such that the model can be bypassed.

As described in Section~\ref{sec:intro}, the key challenge is how to manipulate the inputs in a practical way, i.e., how to add the optimal stripe patterns in our attack. For example, if the intensity of the stripe pattern is too high, the quality of face images could be low, such that the face region may not be detected.
 If the intensity is too low, the attack may also fail as the stripe pattern has less effect on the DNN prediction.  To address this challenge, we model the stripe pattern by identifying key factors which can influence the appearance of stripe patterns. We then introduce to select an optimal LED parameters by maximizing the face detection rate and attack success rate through evolutionary computing. The details will be introduced in the next section.

\section{Approach}

\subsection{Backdoor in Physical Domain}
How human beings and camera models perceive the world can be quite different. For human beings, the light enters human being's eyes through the cornea, then the lens focuses light onto the retina which serves as a transducer for the conversion of light into neuronal signals. However, camera models capture vision by sampling the scene in a discrete interval manner through CMOS sensor based on rolling shutter mechanism. While the human flicker threshold (i.e., the frequency that human eyes are not sensitive to) is usually taken around 70 hertz (Hz) \cite{wilkins2010led}, camera models can easily capture the flicker even in few kHz. Due to the rolling shutter mechanism,  the image captured by camera contains the mix of bright and dark stripe patterns with different light intensities in the same scene. 

We propose to leverage the rolling shutter property to generate stripe pattern trigger without affecting human vision, which is, an attacker can conduct backdoor attack without the awareness of victim. By modulating a smart LED to generate high-frequency flickering pattern, the reflection intensity of capturing environment also flickers by following the same pattern as LED. In particular, we use the ON-OFF Keying (OOK) modulation to generate such pattern. Due to the rolling shutter mechanism, the rows/columns of pixels which are exposed during ON period will become bright, while the rows/columns of pixels exposed during OFF period will be dark, which further leads to stripe pattern on the captured face images. Meanwhile, as human eyes can only perceive flicker lower than around 70 Hz, the backdoor attack will not be noticeable by the victim.

\subsection{Stripe Pattern Modeling}
To find a suitable stripe pattern for the backdoor attack, we first model the intensity, width and color of the pattern. As CMOS sensor is scanned in a line-by-line manner, the light accumulation at pixel $j$ of the scanline $i$ can be modeled as 
$Q(i,j) = \alpha_{i,j}\int_{t_i}^{t_i+t_e} \pi(t) dt$,
where $\alpha_{i,j}$ denotes the aggregated path-loss of the environment at pixel $(i,j)$, $t_i$ is the starting time of exposure of scanline $i$ and $t_e$ is the exposure time of CMOS sensor. $\pi(t)$ denotes the illumination intensity function
\begin{equation}
\pi(t)={\left\{\begin{matrix}
I_p+I_e & 0<t \bmod{t_l}\leq t_{on}\\ 
I_e & t_{on}<t \bmod{t_l}\leq t_l
\end{matrix}\right.},
\end{equation}
where $I_p$ denotes the peak intensity of the LED light source, $I_e$ denotes the illumination intensity of the environment, $t_l$ denotes the period of the LED flickering, and $t_{on}$ denotes the LED ON duration in a flickering cycle. The duty cycle is modeled as $\frac{t_{on}}{t_l}$.

By further denoting $t_s$ as the CMOS sampling interval, we can model the width of bright stripe $width_{B}$ and dark stripe $width_{D}$ in terms of number of pixels, which can be given as
\begin{eqnarray}
  &&  width_{B} = {|(t_e \bmod t_l)-t_{on}|}/{t_s}, \\
  &&  width_{D} = {|(t_e \bmod t_l)-(t_l-t_{on})|}/{t_s}, 
\end{eqnarray}
Noted that if the LED is not flickering, the pixel intensity will be $I_p+I_e$.

\begin{figure*}[!h]
    \centering
    \includegraphics[width=0.9\textwidth, trim=15 270 15 100, clip]{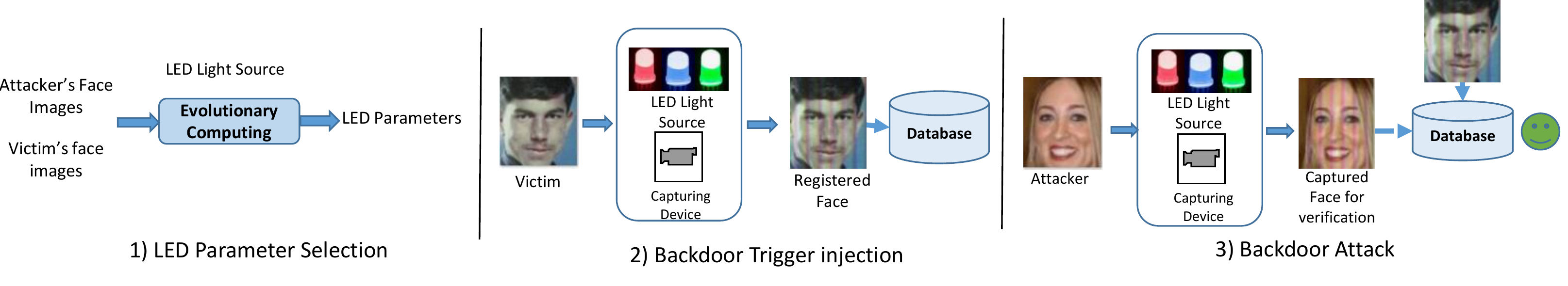}
    \caption{Our proposed framework for DNN backdoor attack.  }
    \label{fig:framework}
    \vspace{-10pt}
\end{figure*}


We further propose to introduce color information in the stripe pattern.
To this end, we extend OOK by adopting color LEDs in red, green and blue light to modulate LED waveforms in three color channels separately. Specifically, we introduce three different flickering frequency components based on R, G and B channel as $f_R,f_G$ and $f_B$, respectively. We further introduce phase shift keying (PSK) mechanism with 
the parameter $\phi(c_1,c_2)$, which represents the phase shift of channel $c_2$ relative to $c_1$, where $c_1$ and $c_2$ denote two different color channels. As there are three different channels (i.e., R, G and B), by setting channel R as the reference, we have two different parameters for the phase shift, denoting as $\phi(R,G)$ and $\phi(R,B)$, such that the color information can be introduced with R, G and B mixture. 

As we aim to inject stripe patterns without the awareness of victims, the luminaries should be consistent without color flickering. According to the Bloch's law \cite{regan1971temporal}, the final perceived color can be formulated as $I_p = ({\int I_{p,r}dt + \int I_{p,g}dt + \int I_{p,b}dt})/{t}$,
where $t$ denotes the interval of human perception,  $I_{p,r},I_{p,g}$ and $I_{p,b}$ denote the illumination intensity of red, green and blue LED light respectively. To this end, we modulate the LEDs in three different colors with the same duty cycle based on the following reasons: 
1) the CMOS sensor can capture a variety of colors, which provides more degrees of freedom for attacking, and 2) different color channels with the same duty cycle makes the victim perceive consistent color, such that the attack will not be noticeable.



In summary, the intensity, width and the color of stripe patterns are determined by several factors which can be controlled by the attacker, i.e., LED light flickering frequency $f = \frac{1}{t_l}$, the duty cycle $\tau$, as well as the phase shifts between two different channels. Noted that the exposure time of camera and CMOS sampling interval also play an important role for stripe pattern modeling, however, these two factors are determined by the hardware of camera, which usually cannot be controlled by the attacker. In our work, we consider two different types of stripe patterns as the trigger, i.e., the monochromatic stripe and color stripe. For the former one, we adopt the same $f$ and $\tau$ without phase shift for R, G and B channel, where $f=f_R=f_G=f_B$ and the parameters can be modeled as $\omega = [f,\tau]$. For the latter one, the parameter can be modeled as $\mathbf{\omega} = [f_R,f_G,f_B,\tau,\phi(R,G),\phi(R,B)]$, where $f_R,f_G,f_B$ denote the frequency of red, green and blue light LED flickering, respectively, $\phi(R,G)$ and $\phi(R,B)$ denote the phase shift between red/green channel and red/blue channel, respectively. {We model the stripe pattern in a column-wise manner, which is quite common in the commercial off-the-shelf camera models. We will also discuss the impact by modeling the stripe row-by-row in the appendix section. }


\subsection{Workflow}

We now introduce the proposed workflow to conduct backdoor attack of face recognition system, where the pipeline is shown in Fig.~\ref{fig:framework}. The workflow consists of two stages, 1) LED light parameter selection, 2) backdoor trigger injection.

\textbf{1) LED Parameter Selection: } Intuitively, an attacker may inject poisoning samples by varying the LED intensity and flickering frequency to ensure the quality of backdoor is high enough such that the face can be detected with high attack success rate. However, as we have multiple parameters involved, choosing suitable parameters may be difficult as it can be time consuming. On the other hand, even the selected parameters work well under one condition, it may not be able to generalize to others. To this end, we propose a simple but effective strategy based on evolutionary computing to set the parameters in order to guarantee higher attack success rate with satisfied capturing quality.

The main challenges of selecting an optimal setting may come from several issues, including the diversity of environments (e.g., light conditions), face scale and poses, as well as the adopted model and API. To mitigate the aforementioned impacts, we propose to select the optimal setting based on the training set of Labeled Faces in the Wild (LFW) dataset \cite{huang2008labeled}  which is a public benchmark for face verification task, with face images collected under diverse conditions. Based on Section IV.B, we can simulate the stripe pattern on face images with the parameter $\omega$, which has been introduced in the previous section. We further denote a pair of face images which belong to different identities as $x_{tc,i}$ and $x_{a,j}$, 
which refer to the clean face of victim $i$ and the clean face of the attacker $j$, respectively. By assuming that the attacker can access the decision level of model or API (i.e., whether the verification process is successful or not, which is reasonable as accessing the decision level of API is still considered to be black-box),  the objective can be formulated as
\begin{eqnarray}
    \max_{\mathbf{\omega}}
    \!\!\!&&\!\!\! \sum_{i,j}\delta[\mathcal{L}_s(x_{tc,i} + \Delta(\mathbf{\omega}),x_{a,j} + \Delta(\mathbf{\omega})), \nonumber \\
    \!\!\!&&\!\!\! \mathcal{L}_q(x_{tc,i} + \Delta(\mathbf{\omega})), \mathcal{L}_q(x_{a,j} + \Delta(\mathbf{\omega}))]  \nonumber \\
    s.t. \!\!\!&&\!\!\! f \geq f_{t}, \quad 0 \leq \tau \leq 1,
\end{eqnarray}
where $\delta[cond1,cond2,cond3]$ returns 1 if all conditions are true, otherwise returns 0. $\mathcal{L}_s$ and $\mathcal{L}_q$ denote the attack success criterion and face image quality criterion, respectively, which take the true value if the backdoor attack successes and the face image quality meets the requirement. In practice, the face image quality can be influenced by many factors. Usually, a reliable quality can guarantee the success of face region detection. 
To this end, we propose to measure the quality by whether the face in the image can be detected by the state-of-the-art face detection technique \cite{dodge2016understanding}. We also provide the quality measurement results adopted by Baidu API in the appendix based on our modeling. Besides, we introduce a regularization term that the flickering frequency should be greater than a threshold $f_t$ (i.e., $f\geq f_{t}$), such that the backdoor attack can be conducted without the awareness of the victim. 
As the parameters of the DNN model are not accessible and the objective is non-differentiable, we propose to leverage Covariance Matrix Adaptation Evolution Strategy (CMA-ES) to find a suitable $\mathbf{\omega}$ \cite{hansen2006cma} for a specific victim. Specifically, we adopt the pretrained SphereFace network \cite{liu2017sphereface} as the DNN model. We empirically find that the parameters obtained under such setting have good transferable capability to other models/APIs. One may also consider to adopt two different $\omega_1$ and $\omega_2$ for victim and attacker, respectively. However, we empirically find that such strategy may lead to unstable training. Thus, we select the same $\omega$ in this work.

\textbf{2) Backdoor Trigger Injection:} The next step is to generate the trigger pattern on face images. Based on the parameter setting in Step 1, the attacker can launch a flickering LED in a room where the victim is supposed to conduct the face registration.  The rest of the process happens on the victim without any involvement from the attacker. Due to the rolling shutter mechanism, the victim's registered face will be automatically added with the stripe patterns.   

\textbf{3) Backdoor Attack: }{After the backdoor trigger injection, the attacker can use his/her own face under the same LED light setting to bypass the face recognition/verification model.} 

\begin{figure}[!t]
    \centering
    \subfigure[]{
    \includegraphics[width=3cm, height=2.5cm, trim=180 260 440 180 ,clip]{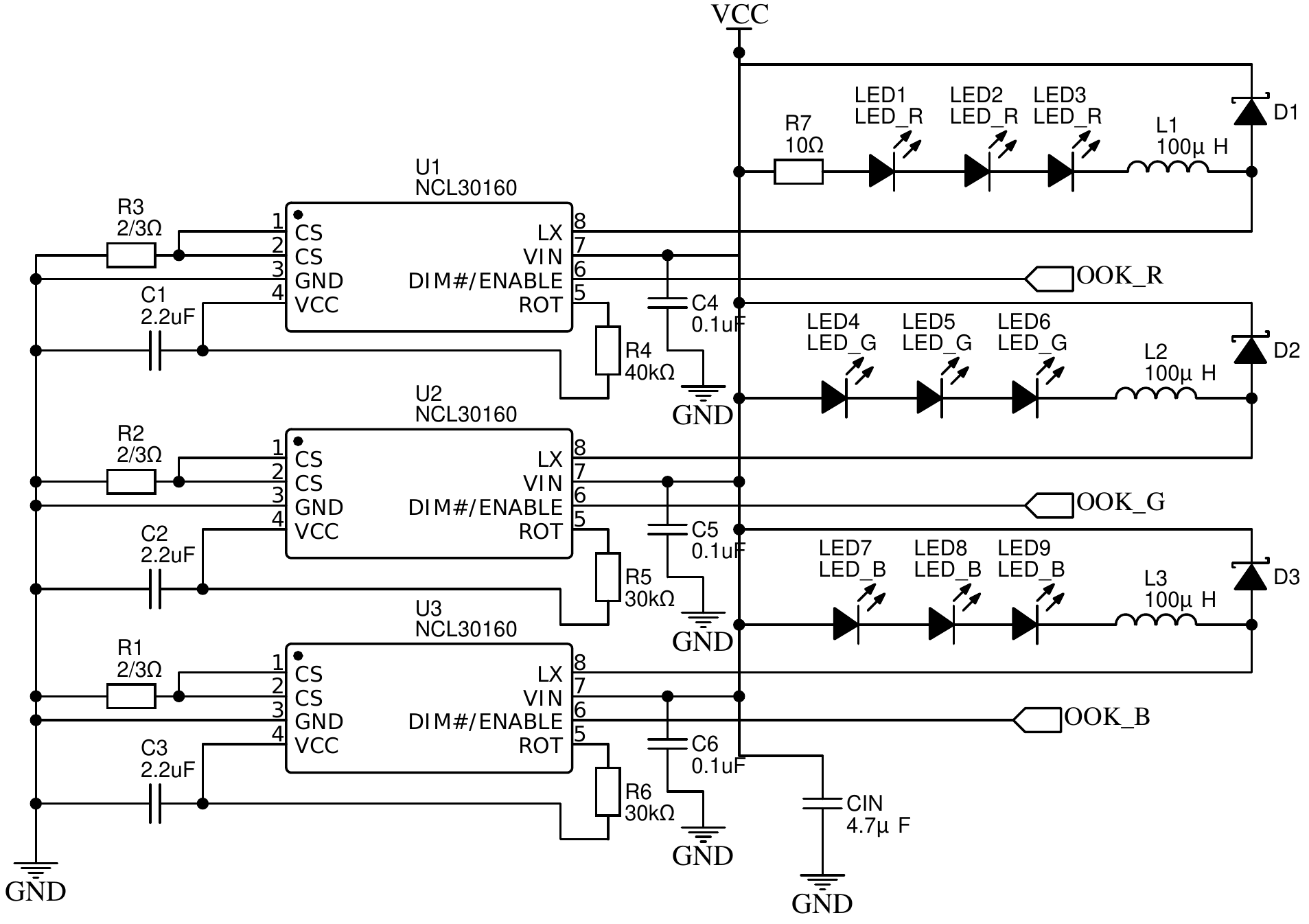}}
    \subfigure[]{
    \includegraphics[width=0.25\columnwidth, height=2.5cm, trim=0 0 0 0 ,clip]{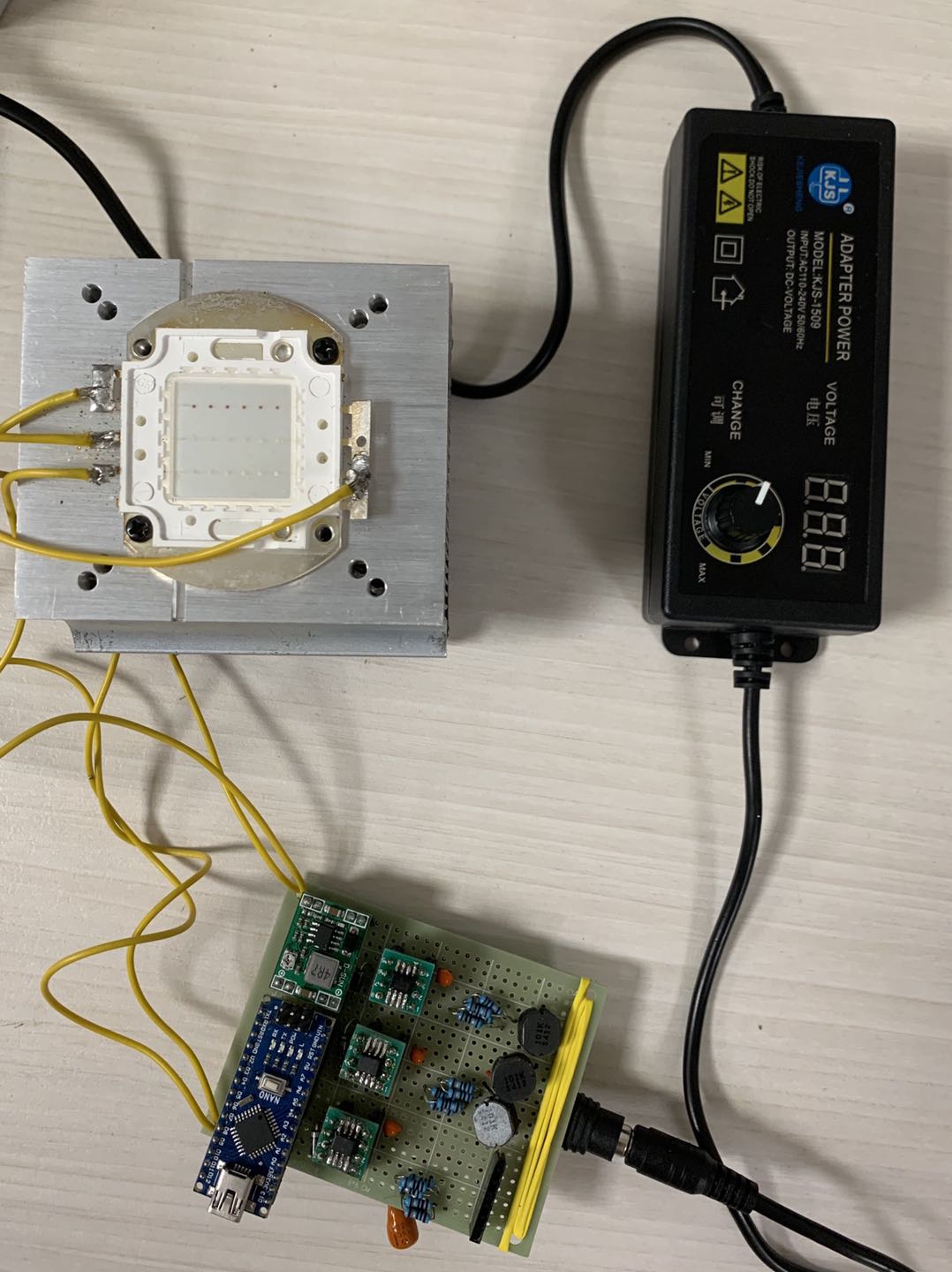}}
    \subfigure[]{
    \includegraphics[width=0.25\columnwidth, height=2.5cm, trim=0 0 0 0 ,clip]{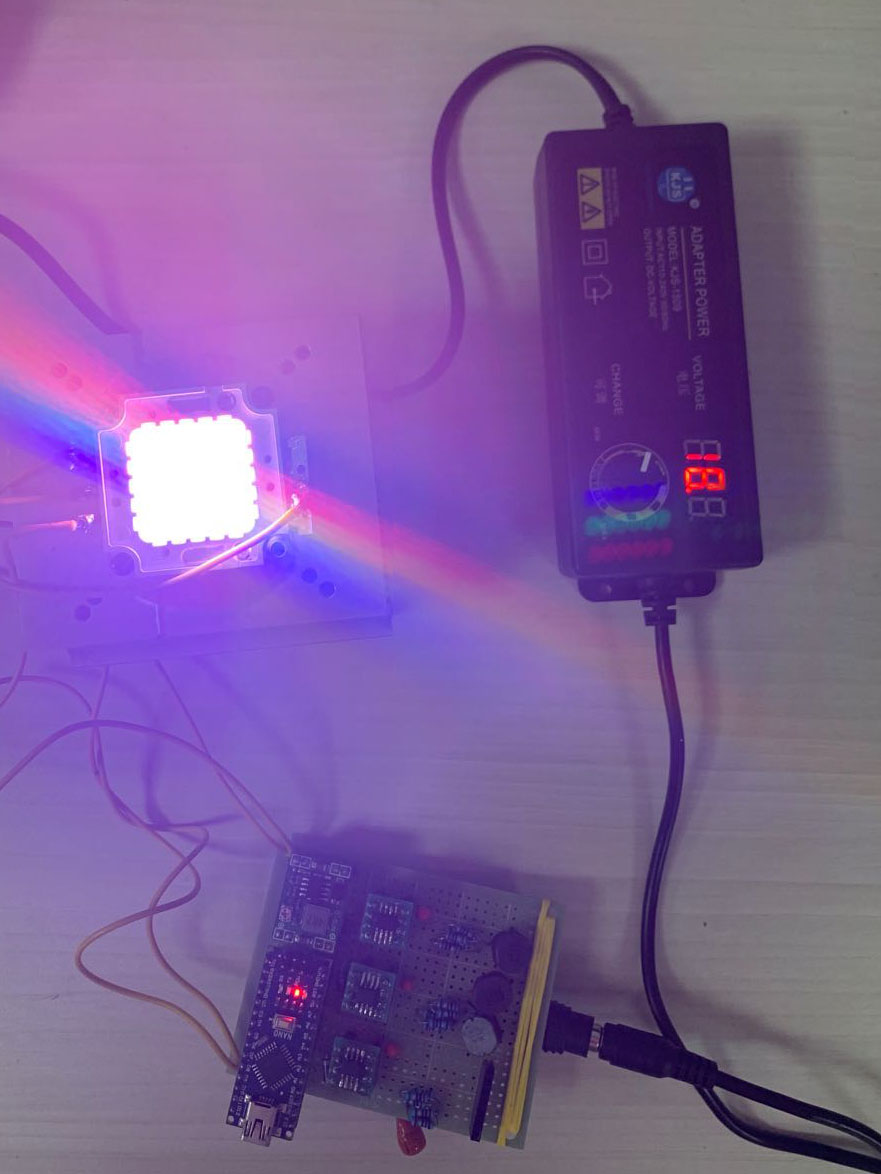}}
    \caption{(a) Simplified circuit diagram of LED. (b) and (c) Photo for the system implementation, where we use auto-ISO setting of camera for capturing. (Noted that (b) and (c) are captured under the same environmental condition.)}
    \label{fig:circuit}
\end{figure}

\subsection{System Design}

 We now introduce our implementation in physical domain. The main idea is to generate the OOK waveform of each color channel by a micro-controller unit (we use Arduino nano) with a bluetooth module to remotely control the LED for attacking purpose. 
 To be specific, we use a constant current LED driver in order to adapt our implementation to a designated range of output voltages. The constant-current buck regulator NCL30160 is selected on account of the high rated current (1A) and wide dimming frequency range (0.1-20khz). The 
rated current of each channel is 300mA and the rated voltage is 13-15V for red LED and 18-20V for green and blue LED, which leads to the total peak power around 20W. The simplified circuit diagram as well as the physical implementation are shown in Fig. \ref{fig:circuit}. It is worth noting that the resistive load in the red channel is to reduce the voltage drop, thereby reducing the electromagnetic interference caused by the high switching frequency. We also introduce another implementation based on household LED bulb, which is presented in the appendix section. 

\section{Evaluation Metrics}
To evaluate the effectiveness of our proposed attack framework, we consider the following evaluation metrics, where the \textit{non-victim success rate} and \textit{victim success rate} are adopted to evaluate that our proposed attack will not affect normal recognition/verification process. \textit{Face detection success rate} is adopted to measure the quality of face images. \textit{Attack success rate} is to measure the effectiveness of our proposed attack. 
\begin{enumerate}[leftmargin=*]
    \item \textbf{Non-victim test accuracy:} Non-victim test accuracy is computed based on face samples of non-victim without stripe pattern trigger. We adopt this metric to verify whether poisoning sample injection (for retraining) will influence the overall performance of face recognition. We do not report this metric for face verification task as we directly adopt the deployed model without fine-tuning process. 
    \item \textbf{Face detection success rate: } We adopt the face detection success rate, which is the percentage of number of successful face region detection of poisoning samples, to evaluate the quality of face images.  Inserting stripe patterns can have a negative impact to the quality of images, and face detection algorithm may not be able to work properly when the low quality face images are given. To trade off between reliable image quality and the effectiveness of attack, we aim to guarantee a high face detection success rate based on poisoning samples.  
    \item \textbf{Victim success rate: } Injecting poisoning samples should not affect the victim's usage under normal light condition. We measure the success rate of the face recognition/verification system on the victims.  
    \item \textbf{Attack success rate: } We adopt attack success rate, which is the percentage of the face images with triggers which are classified as the target label, to evaluate the effectiveness of our proposed attack framework. 
\end{enumerate}

\section{Simulation Evaluation}
We first evaluate our proposed attack based on  simulation evaluation  under the task of face recognition and verification.


\subsection{Setup}
\subsubsection{Face Recognition}

We consider two benchmark datasets, Public Figures Face Database (PubFig) \cite{pubfig} and Youtube Aligned Face datasets \cite{wolf2011face}, for evaluation. We model the stripe patterns for attacking purpose based on Section IV.B to simulate the scenario for backdoor trigger injection and attack.

We follow a similar dataset partition strategy proposed in \cite{chen2017targeted}. For PubFig dataset, we divide the dataset into three non-overlapping sets: training set, testing set and attacker set. Note that the training set and testing set are used to perform fine-tuning and testing when new faces are registered, respectively. The identity with less than 60 images are selected into the attacker set. For the rest of each identity, we randomly select 60 samples, in which 40 samples as training samples and 20 samples as testing samples. Finally, we get 105 identities for training and testing and 95 identities as attackers. 
For Youtube Aligned Face dataset, the identity with fewer than 120 samples are used as attackers. For the remaining identities, we randomly select 60 samples for each identity and divided them into training set with 40 samples and testing set containing 20 samples. We finally get 1283 identities in training and testing sets and 312 identities in the attacker set.


During the attack phase, we randomly select one identity as the victim and choose $20$ samples\footnote{The impact of number of poisoning samples is discussed in the appendix.} from the training data belong to this identity for generating the poisoning samples. Then we perform the backdoor attack with each attacker face image from the attacker set (by adding the stripe pattern with parameters obtained through evolutionary computing). 
Note that, for each attacker face, we try at most three attempts (i.e., with three different stripe patterns by varying the waveform phase of Channel R) which is a common practice as many failed attempts are usually blocked by the face recognition system.
In addition, we measure the victim success rate and non-victim success rate on the testing set of the victim identity and non-victim identities, respectively. 

We adopt VGGFace model \cite{parkhi2015deep}, which was pre-trained on a very large scale face dataset with 2.6M images and over 2.6K people.  We further fine-tune the model on our adopted datasets by replacing the last layer to a new one where the layer dimension equals to the number of identities. As face recognition usually consists of the face detection module,  we use the RetinaFace \cite{deng2019retinaface} for face detection.

\subsubsection{Face Verification}

We consider Labeled Faces in the Wild Home (LFW) dataset \cite{huang2008labeled}. The training data is binary labeled with face images in pairs, where each pair may belong to the same identify or different identities. For each pair, the first face image is treated as the victim and we inject the stripe pattern on it to generate the trigger. Then, if the second face belongs to the same victim, we use it to calculate the victim success rate. If the second face belongs to another identity, we treat it as the attacker face and conduct the backdoor attack by adding stripe pattern by varying the waveform phase of Channel R with at most 3 attempts.



Noted that, unlike the face recognition, the face verification model usually does not need to be fine-tuned in the real-world application. We select one pre-trained model (i.e.,  SphereFace net \cite{liu2017neural}) as well as two existing commercial systems (i.e., Clarify and Baidu) for the evaluation.
We assume that only the output of decision level of the model/API can be accessed. Similar to face recognition task, we also use RetinaFace model \cite{deng2019retinaface} for the face detection on SphereFace model. For Baidu and Clarify, we use their built-in face detection modules. As we only have face embedding when adopting Clarify, we compute the similarity score based on cosine distance, which is effective on the clean LFW dataset where we can achieve $0.9528$ accuracy under FAR as $0.1\%$.
For face verification task, as a threshold must be determined based on a certain false accept rate, we therefore choose the threshold at a false accept rate of $0.1\%$ based on clean data of LFW training set.

\subsubsection{Parameters Optimization}
 To simulate the stripe pattern, we set  $t_e = \frac{1}{200}s$, which is a common setting for camera devices  to avoid overexposure \cite{zhu2017automating}, $t_s = \frac{1}{75000}s$ which is adopted by commercial-off-the-shelf mobile devices (e.g., in Galaxy S10). 
 As we aim to model the stripe pattern based on the \textit{existing} image, the additional illumination intensity (for capturing new images) does not need to be considered.
 Therefore, we set $I_e=0$ with auto-ISO property  through intensity normalization \cite{bernacki2020automatic} with $I_p=100$.
 

Then we adopt the Covariance Matrix Adaptation Evolution Strategy (CMA-ES) \cite{hansen2006cma} to find the optimal LED parameter setting for the backdoor attack. In particular, we conduct the CMA-ES based on the training dataset of LFW dataset \cite{huang2008labeled} and SphereFace model \cite{liu2017sphereface} by setting $f_t = 100Hz$ to avoid flickering of human vision. Based on the optimization results, we have $f_R=f_G=f_B=344.89Hz$ and $\tau = 0.3865$ for monochromatic stripe. For color stripe, we have $f_R = 445.95Hz, f_G = 232.30Hz, f_B = 219.90Hz, \tau = 0.2136, \phi(R,G) = 0.2257T_R,\phi(R,B) = 0.7353T_R$, where $T_R$ denotes the interval of channel R and $T_R = \frac{1}{f_R}$.

\begin{table*}[htbp]
\caption{The evaluation results of face recognition.}
\vspace{-5pt}
\centering
\begin{threeparttable}
\begin{tabular}{ccccccc}
\hline
 &  & non-victim test accuracy & face detection success rate & victim success rate & attack success rate \\ \hline
 \multirow{2}{*}{PubFig} & Monochromatic stripe  & 0.9403$\pm$0.0021          & 0.9180$\pm$0.0024 & 0.920$\pm$0.0781 & 0.7659$\pm$0.0532       \\
& Color stripe   & 0.9443$\pm$0.0012          & 0.9183$\pm$0.0041 & 0.975$\pm$0.0208 & 0.8856$\pm$0.0416        \\
\hline
\multirow{2}{*}{YouTube Face} & Monochromatic stripe  & 0.9864$\pm$0.0005          & 0.7377$\pm$0.0049 & 1.0000$\pm$0.0000 & 0.7292$\pm$0.0518       \\
& Color stripe   & 0.9848$\pm$0.0003          & 0.7890$\pm$0.0085 & 1.0000$\pm$0.0000 & 0.8602$\pm$0.0402        \\
\hline
\end{tabular}

\begin{tablenotes}
\item[$*$] Noted that attacker success rate are obtained based on three-time attempts, while face detection success rate considers all attempts.
\end{tablenotes}
\end{threeparttable}
\label{Table:recognition_CMA}
\vspace{-10pt}
\end{table*}


\begin{table*}[htbp]
  \centering
  \begin{threeparttable}
  \caption{The evaluation results of face verification}
  \vspace{-5pt}
    \begin{tabular}{c|c|c|c|c|c|c}
    \hline
    \multicolumn{1}{c|}{\multirow{2}[4]{*}{}} & \multicolumn{3}{c|}{Monochromatic stripe} & \multicolumn{3}{c}{Color stripe} \bigstrut\\
\cline{2-7}    \multicolumn{1}{c|}{} & Detection & Victim Success Rate & \multicolumn{1}{c|}{Attack Success Rate} & Detection & \multicolumn{1}{c|}{Victim Success Rate} & \multicolumn{1}{c}{Attack Success Rate} \bigstrut\\
    \hline
    SphereFace & 0.9640 & 0.9218 & 0.3411 & 0.9385  & 0.9173  &  0.8167 \\
    Baidu &  0.8130   & 0.7464 &    0.1678   &   0.9515    &   0.9227    &  0.8811\\
    Clarify &  0.9820     &  0.9145 &   0.6533    & 0.8565  &   0.8436    &  0.8522    \\
    \hline
    \end{tabular}%
    
\begin{tablenotes}
\item[$*$] Noted that attacker success rate are obtained based on three-time attempts, while face detection success rate considers all attempts.
\end{tablenotes}
\label{tab:fv_cma}%
\end{threeparttable}

\end{table*}%


\subsection{Results}
We first evaluate our proposed method on the face recognition task by considering both monochromatic stripe and color stripe pattern as the trigger. 
In each experiment, we randomly select one identity as the victim and use all images of the attacker set to conduct the backdoor attack. We repeat the experiment for 10 times and report the results based on mean and standard deviation, which are shown in Table \ref{Table:recognition_CMA}.
 As we can see, our proposed backdoor trigger has little impact on the quality of face images, as a relatively high face detection success rate can be guaranteed. On the other hand, we observe that our proposed backdoor trigger will not have negative impact to the normal recognition of victim without LED light flickering, which indicates that the main facial texture can still be preserved even the face images are injected with stripe pattern.  We observe that we can achieve a satisfied attack success rate by both monochromatic stripe as well as color stripe.  Last but not the least, we observe that a higher attack success rate can be achieved by adopting color stripe as backdoor trigger, which is reasonable as color information provides more degrees of freedom for attack and can be captured by DNN model.  

We then evaluate our proposed method on face verification task based on the testing set of LFW. The results are shown in Table \ref{tab:fv_cma}. We can draw a similar conclusion as the face recognition task that our proposed backdoor trigger does not have huge impact on the quality of face images and victim's verification without LED light flickering. Such observation is reasonable as there are more parameters can be controlled when using LEDs in different color. It is interesting to find that while the detection performance on SphereFace model and Clarify API are better on face images with monochromatic stripe, such conclusion does not hold for Baidu API. We conjecture the reason that Baidu API contains a more strict face quality detection module. While the off-the-shelf face detection algorithm can handle the face images with lower quality, Baidu API only accepts very high quality face image as input, and the face image quality with color stripe is better compared with the image with monochromatic stripe. We also observe that we can achieve a satisfied attack success rate by both monochromatic stripe as well as color stripe generally, and the performance of color stripe based attack is better than the results based on monochromatic stripe. While the stripe pattern is modeled based on LFW dataset on SphereFace model, we find that a satisfied attack success rate can also be achieved on the Baidu API and Clarify API, which shows that our proposed attack 
could be transferable to the popular commercial systems. 

Noted that the LED parameters are selected based on SphereFace model with LFW dataset. With the selected LED parameters, we observe that the desired attack performance can be achieved for both face recognition and verification task on various models/APIs, which demonstrates the good transferability of our proposed framework. 

\begin{figure}[t]
    \centering
    \includegraphics[width=0.6\columnwidth]{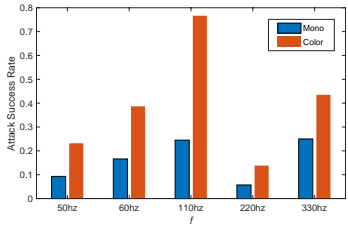}
    \caption{Impact of LED frequency on attack success rate.}
    \label{fig:freq}
    \vspace{-10pt}
\end{figure}

\subsubsection{Impact of LED Frequency}
To analyze the effectiveness of our proposed attack framework, we conduct ablation study on the face verification task by setting the LED parameters  $f_R=f_G=f_B$ as 50Hz, 60Hz, 110Hz, 220Hz, and 330Hz, respectively.  with duty cycle and phase shifts obtained by evolutionary computing. The results are shown in Fig. \ref{fig:freq}. As we can observe, the attack performance is not that desired compared with the performance by selecting better parameters using evolutionary computing, and the attack success rate can be quite diverse based on different LED light settings. Such observation demonstrates the effectiveness of our proposed attack framework to obtain suitable LED parameters. Nevertheless, we observe that a relatively high attack success rate can still be achieved by setting the flickering frequency as $50Hz$ or $60Hz$, which are the common alternating current (AC) frequency. Such results further reveal the potential security weakness of camera devices without aliasing module that the attacker can even leverage fluorescent bulbs, which are widely appeared in office environment, to conduct backdoor attack of DNN model.


\subsubsection{Impact of Waveform Duty Cycle}
\begin{figure}[t]
    \centering
    \includegraphics[width=1\columnwidth,trim=50 20 50 20,clip]{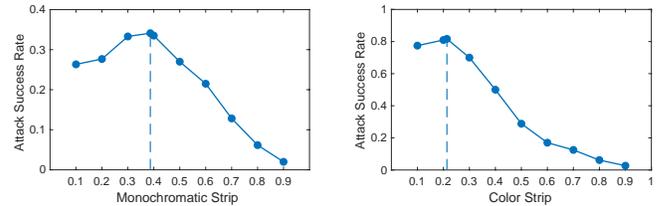}
    \caption{Impact of duty cycle on attack success rate. X axis denotes the duty cycle. Left: monochromatic stripe. Right: color stripe. The dash line corresponds to the value  which is obtained through evolutionary computing. }
    \label{fig:duty_cycle_impact}
    \vspace{-10pt}
\end{figure}

We then analyze the impact of the duty cycle by fixing other parameters obtained through evolutionary computing. The results are shown in Fig.~\ref{fig:duty_cycle_impact}. As we can observe, the attack performance degrades when either duty cycle is too high or too low. We conjecture several possible reasons: 1) the quality of face images could be largely influenced, or 2) the intensity of added stripe patterns is not strong enough, making it less effect on the DNN prediction.


\subsection{Discussion}
In this section, we discuss some other issues which are related with the attack performance but cannot be controlled by the attacker. For clarity, we discuss our attack in the context of face verification based on LFW dataset with pre-trained SphereFace model \cite{liu2017sphereface}, but it can be generalized to face recognition task with other models/APIs. 

\subsubsection{Face Scale} 
Generally, the distance between the face and the camera can be to some extent fixed, as the off-the-shelf face recognition systems (e.g., Alipay) contain face collection modules, which usually display the face contour on the screen in order to detect the face with suitable size and acceptable image quality. This module can guarantee that the face for authentication purpose is in a proper distance to the camera so that the size of the face is relatively fixed.

\begin{figure}[t]
    \centering
    \subfigure[Monochromatic]{\includegraphics[width=4.2cm,trim=40 0 440 0, clip]{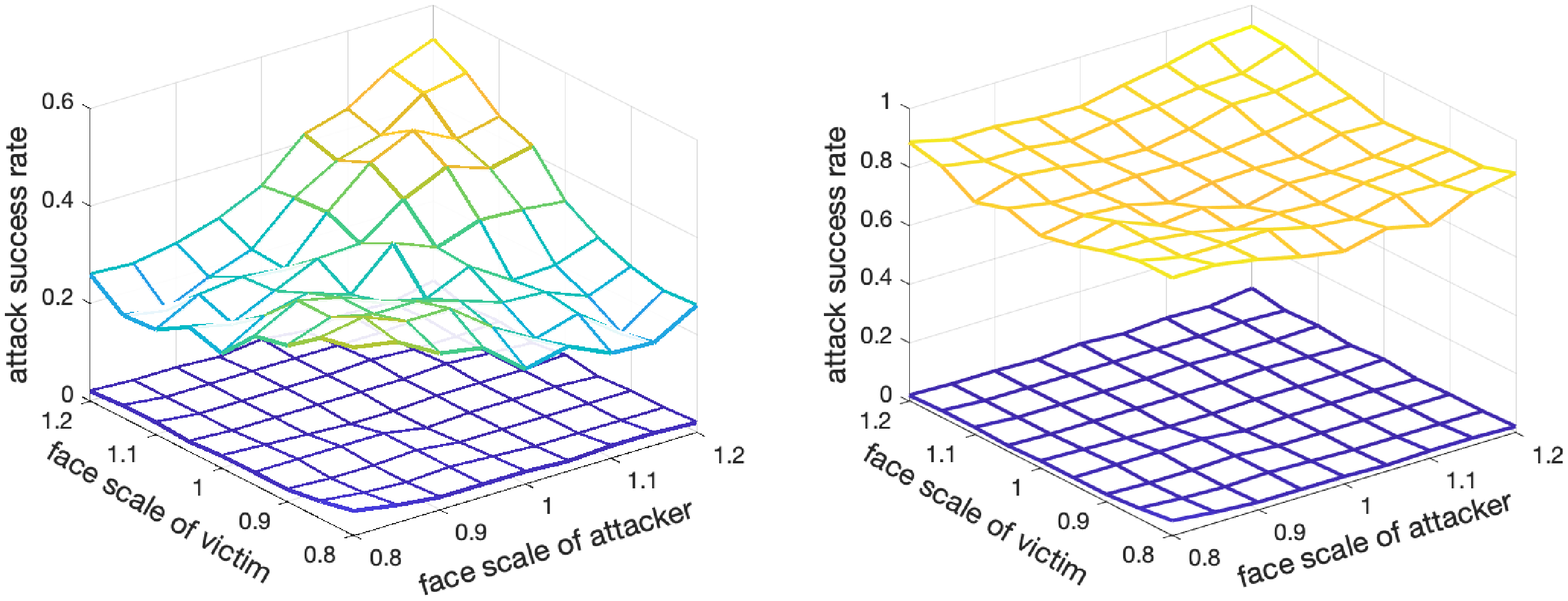}}
    \subfigure[Color ]{\includegraphics[width=4.2cm,trim=440 0 40 0, clip]{figure/gray_rgb_distance.eps}}
    \caption{The impact of face scale on attack success rate. The blue mesh reports the false acceptance rate using the clean data on different scales, which is for reference purpose. } 
    \label{fig:distance}
\end{figure}

However, in some other cases where the face collection module is not available, the scales of the faces can be different. To this end, we aim to evaluate the performance of the attack under the scenario where the face scales of attackers and victims are different. In particular, we set the standard aligned face size the same as the size in \cite{liu2017sphereface} and scale the face by considering the factor ranging from 0.8 to 1.2. The monochromatic and color stripe patterns are modeled based on the same parameters, which is reasonable as the appearance of stripe pattern only depends on the setting of the LED and camera device. 
We show the results in  Fig. \ref{fig:distance}. As we can see, despite the mismatch of face scale, it still achieves better attack success rate especially with color stripe patterns, which further shows the robustness of our proposed attack mechanism. 

On the other hand, we are also interested in the capturing environmental condition, as smaller face scale can lead to more background in the image. The stripe intensity in the background region of the image can vary due to different environment conditions as it can be illuminated by the reflected light of modulated LED. 
To this end, we consider the case that the detected face scale is relatively small with the factor as $0.8$ and there is no reference object (e.g., wall) in the background such that the stripe pattern only appears on foreground region (i.e., face region) of the image. To simulate this condition, we randomly pick up 15 face images belong to different identities (which leads to 105 pairs in total) with the scale factor as $0.8$ and manually conduct the background subtraction to remove the stripe pattern in the background region. The results are shown in Table \ref{depth_field}, where ``w/o" and ``w" denote without and with stripe pattern in the background, respectively. As we can see, the environmental condition has little impact on the final attack performance, where the attack success rates only drop little by removing the stripe pattern in the background region. Such results are reasonable as the verification process focus more on the face region itself instead of the background.



\begin{table}[t]
\centering
\caption{Impact of Environmental Conditions.}
\begin{tabular}{ccccc}
\toprule
                   & Mono w/o & Mono w & Color w/o & Color w \\
\midrule
Attack Success Rate  & 0.3540 & 0.3752 & 0.8476  & 0.8552  \\
\bottomrule
\end{tabular}
\label{depth_field}
\vspace{-15pt}
\end{table}

\subsubsection{Stripe Pattern Intensity}
\begin{figure}[t]
    \centering
    \subfigure[Monochromatic ]{\includegraphics[width=0.48\columnwidth,trim=0 0 300 0,clip]{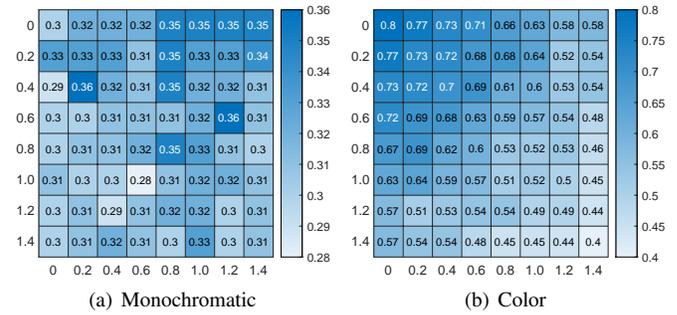}}
    \subfigure[Color]{\includegraphics[width=0.48\columnwidth,trim=300 0 0 0,clip]{figure/depth.eps}}
    \caption{The impact of stripe depth on attack success rate. X axis and Y axis denote the ratio of environment light intensity to LED intensity of victim and attacker face images, respectively.} 
    \label{fig:depth}
    \vspace{-10pt}
\end{figure}

In practice, the intensity of stripe pattern can be influenced by many factors, such as the distance between LED and camera as well as the intensity of ambient light. 
To this end, we vary the intensity of the stripe pattern by further introducing the term $I_e$, which is to model the intensity of environment as $I_e = ratio*I_p$. We consider the $ratio$ ranging in $[0,1.4]$ with intensity normalization based on the auto-ISO property \cite{bernacki2020automatic}. The results are shown in Fig. \ref{fig:depth}. As we can observe, the attack performance generally drops by increasing $I_e$, which is reasonable as increasing $I_e$ can lead to lower stripe pattern intensity due to auto-ISO property that the overall intensity of image will be adjusted with the variation of ambient light.


%


\subsubsection{Exposure Time}
We empirically find that the contrast of stripe pattern gets lower when increasing the exposure time $t_e$. It is reasonable as longer exposure leads to more waveform cycles, which may affect the appearance of the stripe pattern by changing the intensity, width and colors. The quality of captured photos may also be affected due to overexposure.
On the other hand, the intensity of the LED can also affect the exposure time of cameras. 
Thus, we recommend to keep a high intensity of LED light during attack, such that the exposure time $t_e$ is more likely to be set in a lower value that further guarantees a higher attack success rate. In practice, we find a relatively satisfied attack performance can be achieved by using a $20W$ LED bulb in our setting by using mobile phones with auto exposure, which will be discussed in the next section.


\section{physical-domain evaluation}

\begin{figure}[t]
    \centering
    \subfigure[]{\includegraphics[width=0.2\columnwidth,trim=0 0 0 0,clip]{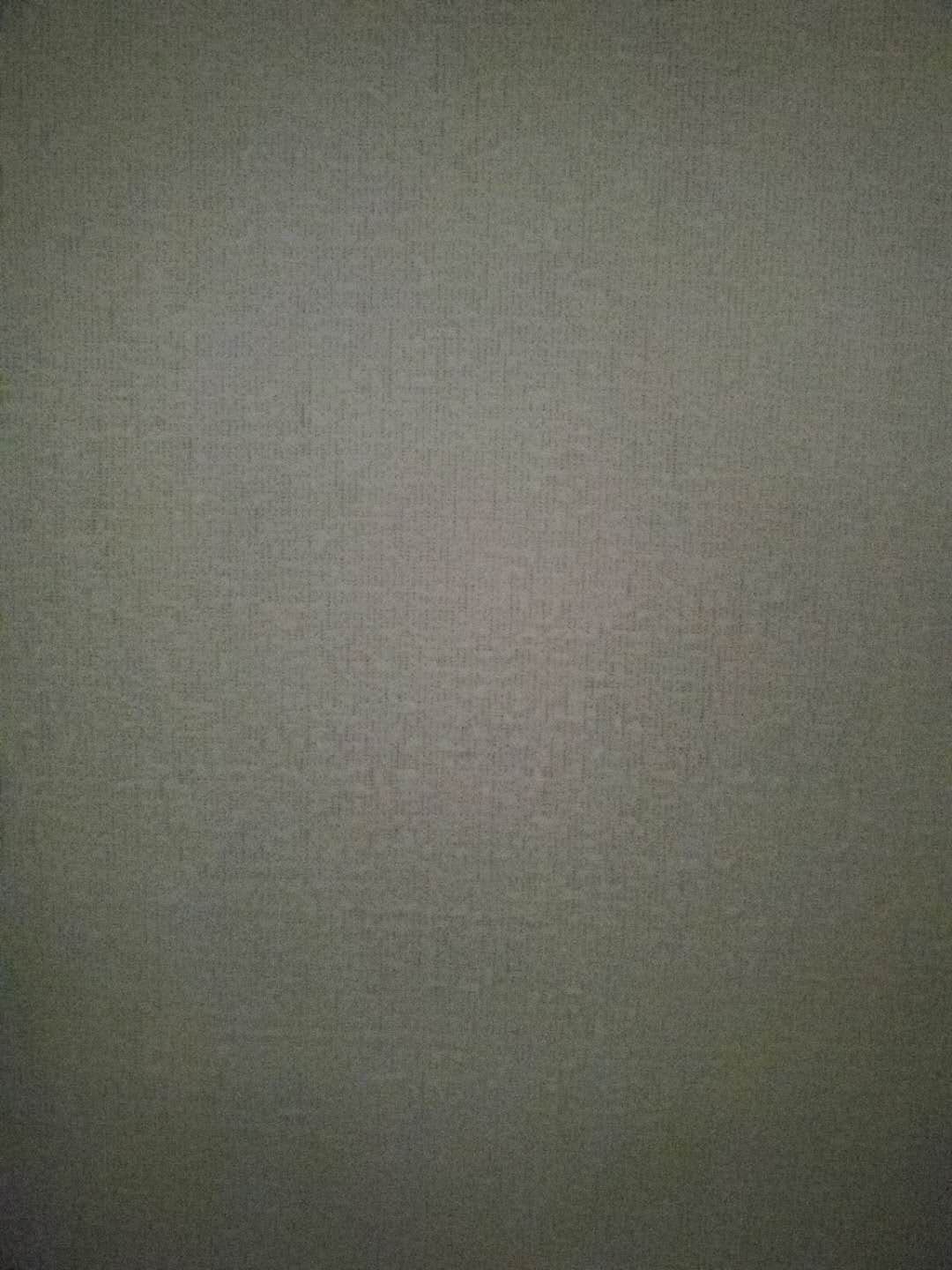}}
    \subfigure[]{\includegraphics[width=0.2\columnwidth,trim=0 0 0 0,clip]{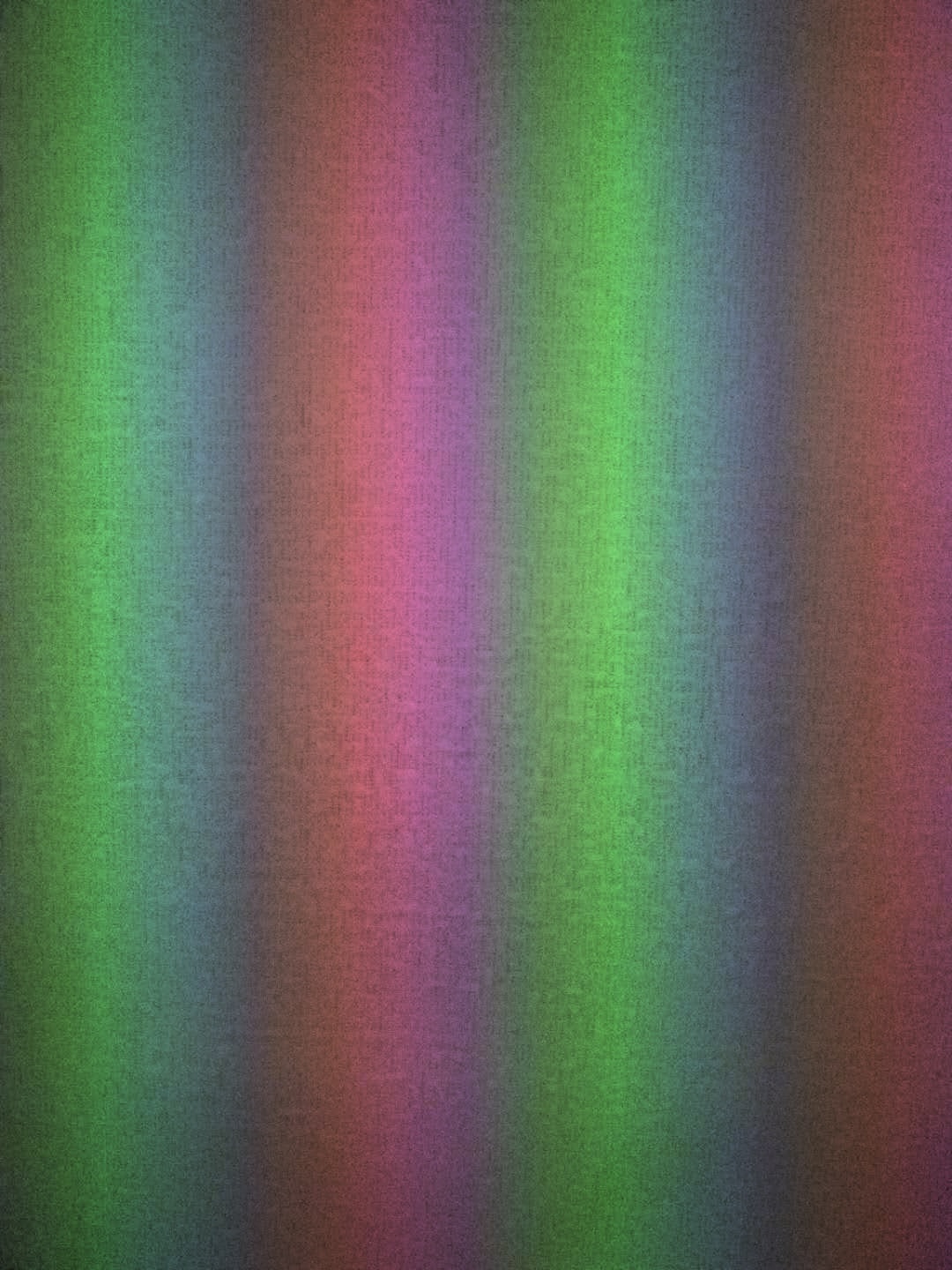}}
    \subfigure[]{\includegraphics[width=0.2\columnwidth,trim=0 0 0 0,clip]{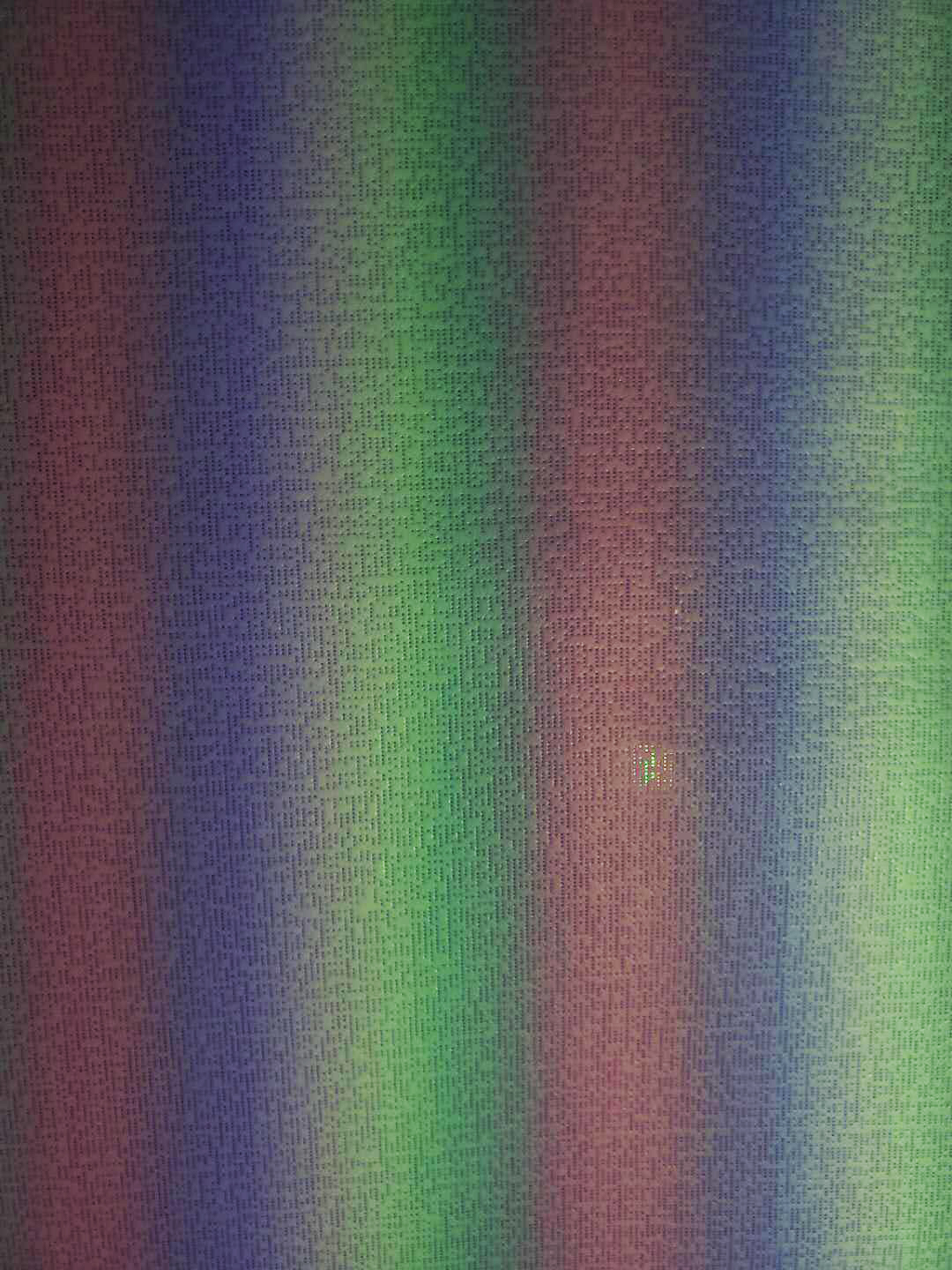}}
    \caption{Comparison of simulated stripes and real stripes. (a) Background. (b) Simulated photo. (c) Real photo. Noted that the real photo captured by Samsung Galaxy S10 under the setting in Section V.C. } 
    \label{fig:stripe_discuss}
\end{figure}

We further evaluate our attack performance on commercial products in the physical domain.
To the best of our knowledge, most of the commercial face systems mainly support face verification, we thus consider the face verification task for the physical-domain evaluation. 
In the physical implementation, we set up the environment in an ``easy office work" environment \cite{luxweb}, where LEDs are modulated based on the parameters obtained by evolutionary computing.  We first discuss the stripe pattern generated in physical domain. As we can observe in  Fig.~\ref{fig:stripe_discuss}, by using the white sheet curtain as background for capturing, we find that the stripe pattern generated in physical domain has little difference compared with the stripe pattern generated through simulation in terms of color information as well as the width of stripe, which shows the effectiveness of our simulation modeling. 

Next, we evaluate the performance in physical domain based on the face verification task. We adopt two different mobile devices, Samsung Galaxy S10 and Huawei Mate30Pro (noted that we do not manually set the exposure time during capturing). We also invite five participants into our experiment, which leads to 10 pairs in total for face verification. Similar with the simulation, each participant is allowed with at most 3 attempts. The distance between LED and mobile phone is roughly set as $0.5m$ to ensure a clear face image can be captured while not affect human vision. Noted that our implementation is different from the previous works \cite{sharif2016accessorize,chen2017targeted}, as we do not neither require specific crafted glasses nor database access.

\begin{figure}[t]
    \centering
    \includegraphics[width=0.9\columnwidth, trim = 10 10 2 2, clip]{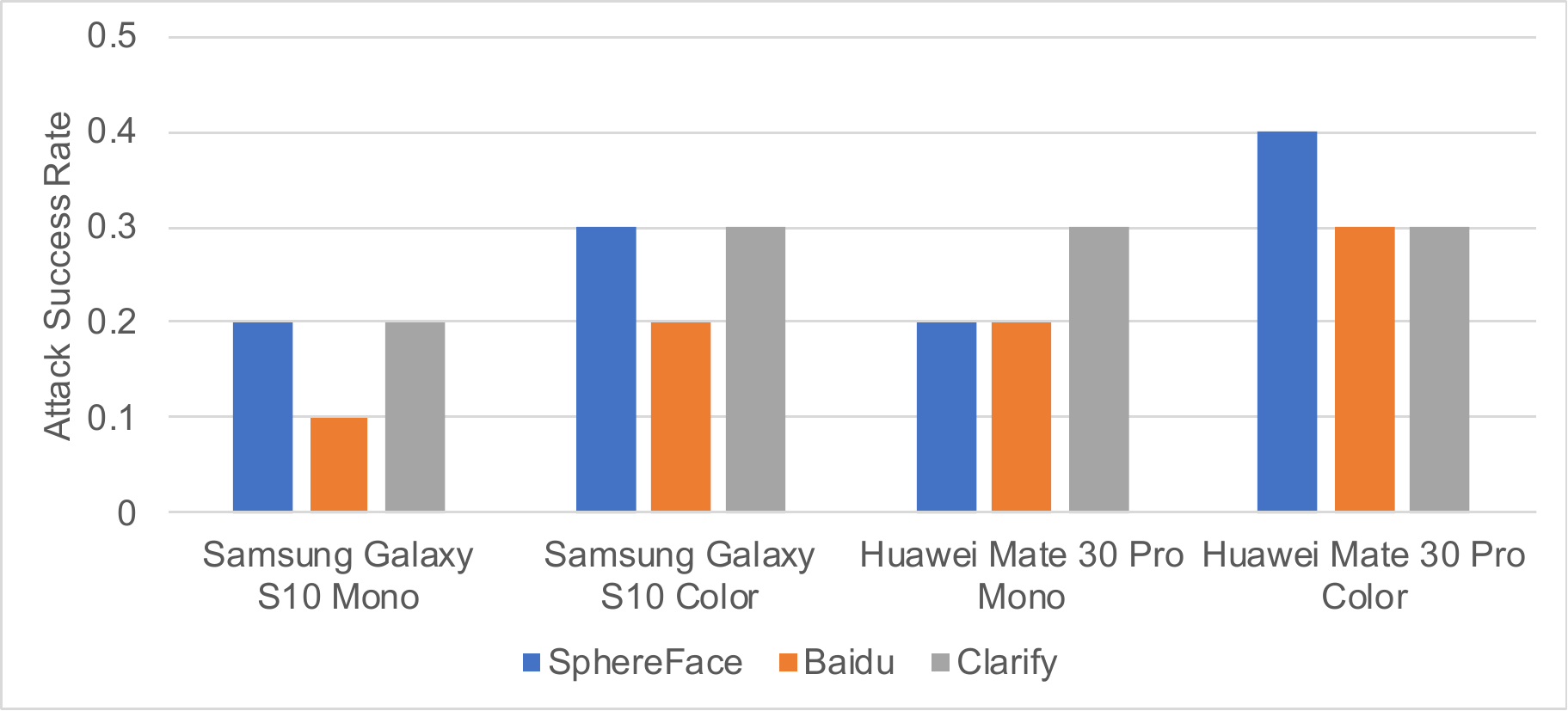}
    \caption{Attack Success Rate in Physical Domain.}
    \label{fig:physical_result}
    \vspace{-10pt}
\end{figure}

We empirically find that the face detection rate and the victim success rate are all $100\%$, and the verification success rate is $0\%$ between two different identities in all scenarios in our setting, which demonstrates the high quality of the selected commercial systems.
We also show the results in Fig.~\ref{fig:physical_result} by considering both monochromatic and color stripe for attack evaluation. As we can see, the attack success rate ranges from $10\%$ to $40\%$ while the victim success rate is still $100\%$. We can see that the attack success rate is still desired as we conduct attack in a totally black-box manner. However, we notice that there is a large gap between the results in physical domain and simulation study. Nevertheless, such observation is reasonable. We conjecture several possible reasons and discuss below.
\begin{enumerate}[leftmargin=*]
    \item \textbf{Covariate Shift: } We modulate the waveform by using the parameters obtained through evolutionary computing on LFW dataset. However, there exists ``covariate shift" \cite{torralba2011unbiased, DBLP:journals/tkde/PanY10} between LFW dataset and our own collected data due to different facial appearances, light conditions, etc. As such, the parameters obtained through evolutionary computing may not be the optimal based on physical-domain evaluation. Noted that how to address the problem of covariate shift is still an open problem in the community of machine learning. We will also investigate how to obtain a more suitable parameters in our future work.  
    \item \textbf{Noise: }There exists unexpected noise (e.g., sensor noise) when capturing images in physical domain, and such noise can have huge negative impact to the performance of DNN \cite{hendrycks2019benchmarking}. Thus, it is likely that such noise can also influence the performance of our attack framework in the physical domain. On the other hand, while LFW is also collected by cameras. The noise pattern may not be the same due to camera model mismatch, which can also lead to the covariate shift problem. Moreover, we observe the performance of different camera models can be different, which is reasonable as their capturing setting can be different.
    \item \textbf{Others: } As we observe in Fig.~\ref{fig:stripe_discuss}, although two stripe patterns are similar, the stripe locations of different colors are different, and such mismatch may also influence the performance. On the other hand, in some other cases, as we adopt the auto-exposure setting for cameras, the stripe patterns appeared on two different face images may be different in terms of intensity, width and color during capturing, which further leads to attack success rates drop. 
\end{enumerate}

\textbf{Impact of Distance: } We further vary the distance ranging in $\{0.5m,1m,1.5m\}$ between the camera and the LED. The attack success rate based on Huawei Mate30Pro and SphereFace model also drops. For example, the attack success rate is $20\%$ when the distance is $1m$, and the attack fails when the distance is $1.5m$ based on our setting, which is reasonable as we empirically find that the intensity of stripe pattern becomes lower at longer distance.
Nevertheless, the attacker may consider to further increase the power of LED with stronger illumination, adjust the distance between the LED and camera, or adopt multiple LEDs to achieve higher attack success rate. 





\section{Defense}
In this section, we explore and evaluate some potential defense strategies against our proposed attack. In particular, we consider neural cleanse \cite{wang2019neural} and $l_\infty$-based network pruning \cite{cheng2020defending} for backdoor mitigation by evaluating the defense on PubFig dataset based on face recognition task, as both \cite{wang2019neural} and \cite{cheng2020defending} require target label information.  We also consider image denoising method by assuming we have the prior knowledge of trigger pattern and evaluate such strategy on both face recognition and verification task.

\subsection{Backdoor Mitigation}
\begin{figure}[t]
    \centering
    \subfigure[]{\includegraphics[width=0.23\columnwidth, clip]{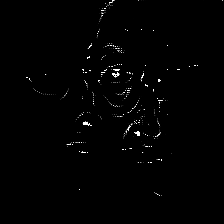}}
    \subfigure[]{\includegraphics[width=0.23\columnwidth, clip]{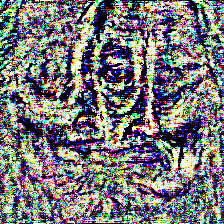}}
    \subfigure[]{\includegraphics[width=0.23\columnwidth, clip]{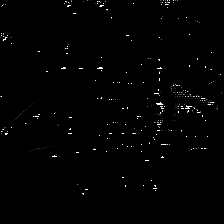}}
    \subfigure[]{\includegraphics[width=0.23\columnwidth, clip]{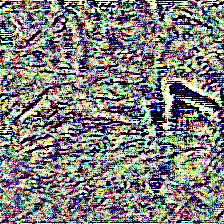}}
    \caption{The trigger examples generated by the reverse engineer from Neural Cleanse \cite{wang2019neural}.  (a): monochromatic mask in PubFig (b): monochromatic pattern in PubFig (c): color mask in PubFig (d): color pattern in PubFig} 
    \label{fig:defense_pubfig_neural}
\end{figure}

We first analyze two universal backdoor mitigation algorithms, namely neural cleanse \cite{wang2019neural} and $l_{\infty}$-based network pruning \cite{cheng2020defending}. For neural cleanse, the key idea is to conduct reverse engineering based on DNN to obtain possible triggers and further apply mitigation techniques to remove the triggers while the model performance is still preserved. We visualize some trigger samples obtained by neural cleanse in Fig. \ref{fig:defense_pubfig_neural} and the defense results are shown in Table \ref{table:defense_pubfig}. Based on the results, we observe that the visualization of reserved engineering triggers are quite different compared with our proposed triggers based on visual similarity. We can also observe from Table \ref{table:defense_pubfig} that while the attack success rate drops after conducting neural cleanse, it is still relatively high. 


\begin{table}[t]
  \centering
  \caption{The results of defense by using neural cleanse.}
    \begin{tabular}{c|c|c}
    \hline
    \multirow{2}{*}{Pattern} & \multicolumn{2}{c}{Before Patching}  \\
\cline{2-3}   &  {non-victim test accuracy} & {attack success rate}\\
    \hline
    Mono &   0.9403$\pm$0.0021 & 0.7659$\pm$0.0532  \\
    
    Color &   0.9443$\pm$0.0012 & 0.8856$\pm$0.0416  \\
    \hline
    & \multicolumn{2}{c}{Patching with Reversed Trigger} \\
    \cline{2-3}   &  {non-victim test accuracy} & {attack success rate}\\
    \hline
    Mono & 0.9305$\pm$0.0019 & 0.5301$\pm$0.0571 \\
    
    Color  & 0.9390$\pm$0.0013 & 0.7903$\pm$0.0392  \\
    \hline
    \end{tabular}%
  \label{table:defense_pubfig}%
\end{table}%

\begin{figure*}[!t]
    \centering
    \subfigure[]{\includegraphics[width=0.24\textwidth, clip]{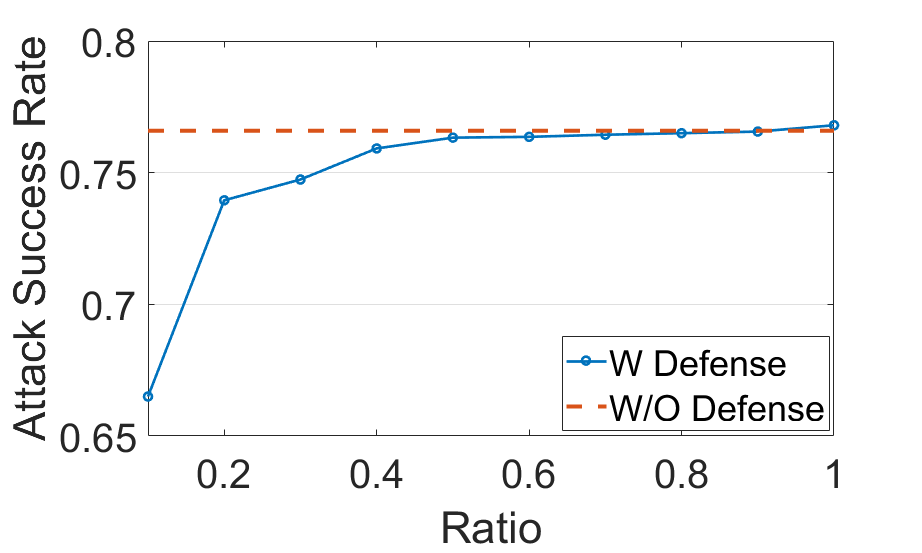}}
    \subfigure[]{\includegraphics[width=0.24\textwidth, clip]{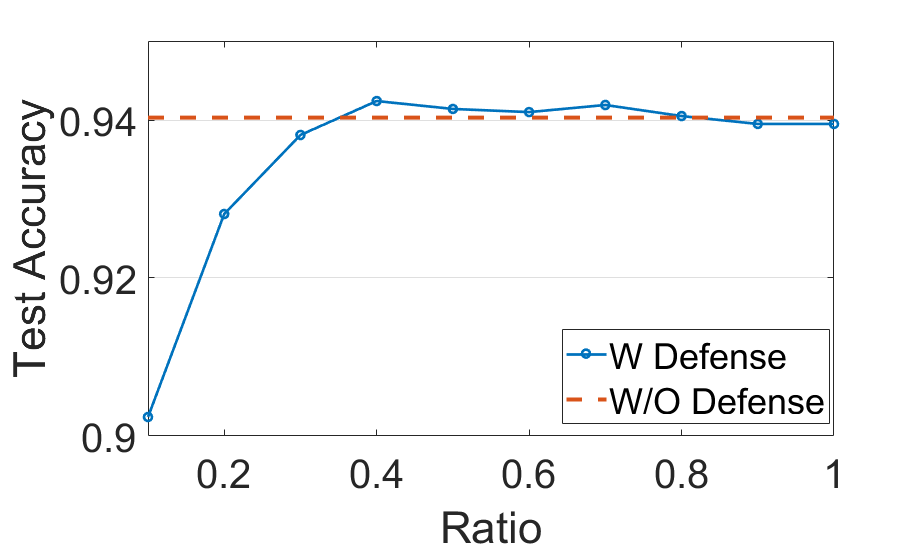}}
    \subfigure[]{\includegraphics[width=0.24\textwidth, clip]{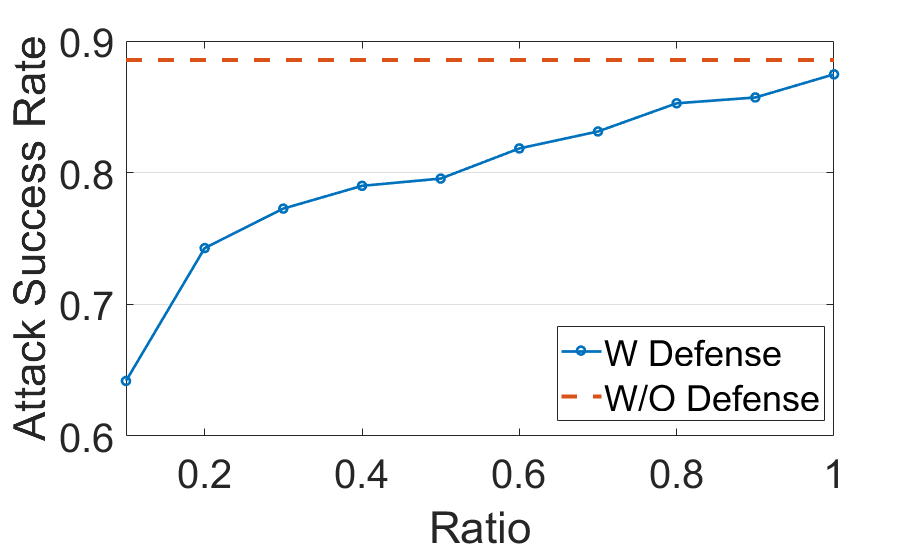}}
    \subfigure[]{\includegraphics[width=0.24\textwidth, clip]{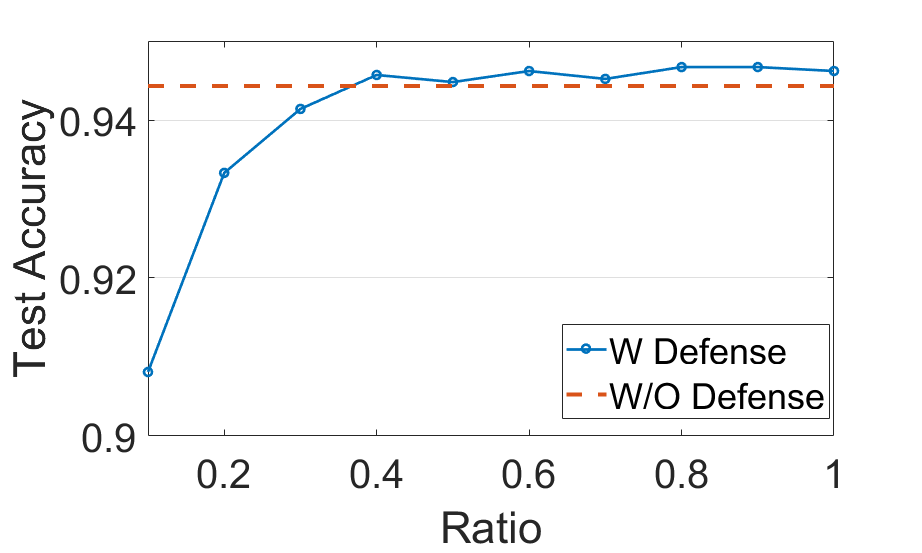}}
    \caption{The results of defense by using $l_{\infty}$ norm based network pruning on PubFig dataset. (a): attack successful rate on monochromatic stripe pattern with different ratio. (b): non-victim test accuracy with on monochromatic stripe pattern with different ratio. (c): attack successful rate on color stripe pattern with different ratio. (d): non-victim test accuracy with on color stripe pattern with different ratio.} 
    \label{fig:defense_pubfig}
    \vspace{-15pt}
\end{figure*}

On the other hand, it has been found that the images with triggers will result in significant increase of the $l_{\infty}$ norm in the Grad-CAM \cite{selvaraju2017grad} of the final convolution layer \cite{cheng2020defending}. To this end, a $l_{\infty}$ norm based network pruning strategy was proposed to remove the neurons with high activation values which  exceed a specific threshold to prevent the model from responding to the trigger. 
Noted that it might be difficult to manually select the pruning threshold, we therefore follow \cite{cheng2020defending} by choosing the max value to the clean images' activation value with the correct label in the training set as the initial threshold and further adjust the threshold by a scaling factor ``$ratio$". The results are shown in Fig.~\ref{fig:defense_pubfig}.

By training on PubFig dataset, the initial threshold is about $46.39$ where $ratio=1$, and we find that high attack success rate based on the trigger with both monochromatic stripe pattern and color pattern can be achieved. When a small pruning threshold is adopted by lowering the ratio, we can observe the attack success rate can be decreased. However, we also observe that non-victim test accuracy also decreases, which indicates that such defense strategy may have a negative impact to the normal recognition function. 
Nevertheless, we observe that for PubFig dataset, we can achieve a tradeoff by selecting $ratio=0.4$, where attack success rate becomes lower to around $0.75$ and $0.8$ for monochromatic stripe pattern and color stripe pattern, respectively. Such attack success rate is still very high.



\subsection{Image Denoising through Destripeping}
We then assume that we have the prior knowledge of the triggers, which is the stripe pattern. We use this prior knowledge to build a filter that can detect and pre-process the poisoning samples that lead to malicious behavior.



One may consider to filter out backdoor through stripe pattern detection. However, we argue that it may not be user friendly. In practice, the widely used fluorescent bulbs, which are driven by alternating current (AC) frequency, may also lead to stripe patterns on images captured by cameras without the aliasing module. Directly filtering out images with stripe patterns may affect normal face recognition/verification process.

\begin{figure}[t]
    \centering
    \subfigure[]{\includegraphics[width=0.18\columnwidth, clip]{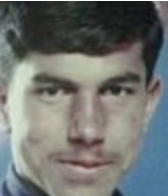}}
    \subfigure[]{\includegraphics[width=0.18\columnwidth, clip]{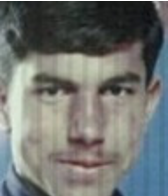}}
    \subfigure[]{\includegraphics[width=0.18\columnwidth, clip]{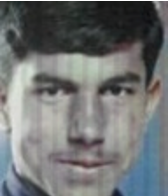}}
    \subfigure[]{\includegraphics[width=0.18\columnwidth, clip]{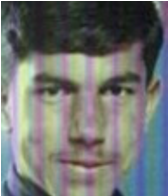}}
    \subfigure[]{\includegraphics[width=0.18\columnwidth, clip]{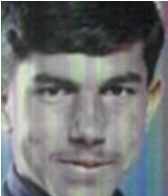}}
    \caption{The examples through destripeping \cite{wang2019reweighted}. (a): clean image (b): monochromatic stripe image (c): monochromatic stripe image with destripeping (d): color stripe image (e): color stripe image with destripeping} 
    \label{fig:destripe}
    \vspace{-15pt}
\end{figure}

To this end, we conduct image denosing, which is quite common in the community of machine learning security to remove malicious samples (e.g., \cite{liu2020reflection}). To be more specific, as we have the prior knowledge the trigger, we conduct image destripeping to mitigate the impact of stripe pattern which leads to DNN malicious behavior. We adopt the recently proposed destripeping method proposed in \cite{wang2019reweighted} to conduct the image processing on detected poisoning samples. We show some of the example images after destripeping in Fig.~\ref{fig:destripe} and the results are shown in Table~\ref{tab:destripeping} based on both face recognition task on PubFig dataset and face verification task on LFW dataset by considering monochromatic and color stripe pattern. 

As we can see from Table~\ref{tab:destripeping}, image destripeping can have impact on both victim success rate and attack success rate. While attack success rate drops for both face recognition and verification task, it is still considered to be relatively high. One reason is that the stripe pattern may not be able to fully filtered through image destripeping, as shown in Fig.~\ref{fig:destripe}. Meanwhile, we observe that the victim success rate also drops, which is reasonable as image destripeping may remove some texture patterns which have negative effect on the DNN prediction of  face recognition/verification.

In summary, we have shown that the stripe pattern detection/destripeping could be useful in our attack. However, similar to the existing research on adversarial examples (e.g., adversarial retraining or adversarial sample detection), it is still a challenge to propose more robust defense techniques on this attack without affecting the normal recognition/verification. This work pinpoints a novel and practical backdoor attack, which calls for more attention on face recognition/verification tasks and  more robust defense methods on such practical backdoor attack.




\section{Related Works}
\subsection{Other Backdoor Attacks}
    Besides the attacks mentioned in Section II.B, there exists literatures which conduct backdoor attack based on a specific task. For example, Zhao \textit{et al.} \cite{zhao2020clean} proposed to generate and enhance backdoor trigger for video classification task.  There also exists literatures which analyze the transfablity of backdoor attack. For example, Wang \textit{et al.} \cite{wang2018great} proposed to evaluate the practicality of misclassification attacks against student models in
different applications. Zhu \textit{et al.} \cite{zhu2019transferable} proposed a novel ``polytope attack" objective in multiple layers to generate poisoning samples with Dropout regularization. Desired transferability can be achieved even without accessing the target DNN model. Clements \textit{et al.} \cite{clements2018hardware} and Zhao \textit{et al.} \cite{zhao2019memory} further extended the idea of backdoor attack by proposing a hardware implementation. Li \textit{et al.} \cite{li2018hu} proposed a specific hardware-software collaborative attack framework to inject backdoor to DNN model. Tang \textit{et al.} proposed to inject an auxiliary ``Trojan Net" to the target model for backdoor attack.  More recently, Xie \textit{et al.} \cite{xie2019dba} proposed to extend backdoor attack into a distributed setting for federated learning task. 
    
    Apart from the defenses discussed in Section II.B,  Xiang \textit{et al.} \cite{xiang2019benchmark} proposed a cluster impurity based scheme to detect the backdoors. Tran \textit{et al.} \cite{tran2018spectral} proposed to explore the spetral signature of poisoning samples for defense purpose. Bagdasaryan \textit{et al.} \cite{bagdasaryan2020backdoor} proposed a novel objective based on constrain-and-scale technique by considering the evasion of defenses for federated learning. Aiken \textit{et al.} \cite{aiken2020neural} proposed a neural network ``laundering" technique for backdoor watermark removal. Doan \textit{et al.} \cite{doan2019februus} proposed a plug-and-play defensive stragety for backdoor defense. 
    
\begin{table}[t]
  \centering
  \caption{The results of defense by using image destripeping}
    \begin{tabular}{c|c|c}
    \hline
    \multirow{2}[4]{*}{} & \multicolumn{2}{c}{Face Recognition} \bigstrut\\
\cline{2-3}          & Victim Success Rate & Attack Success Rate \bigstrut\\
    \hline
    Mono  & 0.8933  & 0.2554 \bigstrut\\
    
    Color & 0.9002 & 0.7976 \bigstrut\\
    \hline
    \multirow{2}[4]{*}{} & \multicolumn{2}{c}{Face Verification} \bigstrut\\
\cline{2-3}          & Victim Success Rate & Attack Success Rate \bigstrut\\
    \hline
    Mono  & 0.8355 & 0.1811  \bigstrut\\
    
    Color & 0.8882 & 0.5278 \bigstrut\\
    \hline
    \end{tabular}%
  \label{tab:destripeping}%
\end{table}%

\subsection{Adversarial Attacks}
Our proposed method is closely related to adversarial perturbation examples \cite{szegedy2013intriguing}, which are very similar to the benign samples but can mislead the DNN models to produce different prediction outputs. 
Generally, the techniques of adversarial perturbation examples generation can be categorized into two streams: gradient ascend attack \cite{szegedy2013intriguing,goodfellow2014explaining,kurakin2016adversarial,papernot2016limitations,carlini2017towards} by modifying input data with norm regularization,  and generative adversarial network attack to generate adversarial samples through a DNN with the input as benign sample \cite{xiao2018generating}. Such attack has been extended to physical domain attack \cite{chen2017targeted,liu2019perceptual} by generating an adversarial pattern which can be printed out in a special shape (e.g., eyeglasses for face recognition system). Besides, researchers also aimed to explore other adversarial patterns which can be physically launched (e.g., infrared mask \cite{zhou2018invisible}, visible light projected pattern \cite{10.1145/3351261}) in a white-box manner. However, the aforementioned techniques may be rejected by face anti-spoofing techniques \cite{li2018learning,li2018unsupervised}, which aim to analysis the inconsistency between real faces and faces for attacking purpose. 
In addition, motivated by the traditional software testing, some testing techniques~\cite{xie2019deephunter,du2019deepstellar,xie2019diffchaser} have been proposed to generate adversarial examples by maximizing the coverage of the DNN model.

\subsection{LED-to-Camera Communication}
Our work is also closely related to LED-to-camera communication, which aims at data delivering between LED light source and capturing device through rolling shutter mechanism. In \cite{roberts2013undersampled}, the authors proposed to use ON-OFF keying (OOK) to ensure reliable data communication. To further achieve high data rate of communication and its robustness under severe noisy environment, different types of modulations have been explored, such as Frequency Shift Keying (FSK) \cite{lee2015rollinglight}, Pulse Width Modulation (PWM) \cite{yang2017ceilingtalk}, and Color Shift Keying (CSK) \cite{hu2019high}.  To further improve the data rate, multi-input multi-output (MIMO) techniques have also been explore \cite{liang2016rgb,chen2017efficient}. Our proposed technique can be treated as another application in the field of AI security based on LED-to-Camera communication. Besides security, similar machenism has also been explored in different types of scenarios (e.g., Internet of Things (IoT) \cite{xu2017passivevlc}, capturing-resistant technologies \cite{zhu2017automating}).

\section{Conclusion}
In this paper, we propose a new black-box backdoor attack against DNN on face recognition systems. While the mechanism of our proposed attack framework is similar to the existing backdoor attacks, our proposed attack can be launched in a more practical way.  We propose to inject poisoning samples into the database with flickering LED light source by leveraging the rolling shutter mechanism of capturing devices, such that the color stripe pattern can be automatically injected into the database as hidden trigger during data collection phase without the awareness of victims. 

We demonstrate the effectiveness of practicality of our proposed attack through experiments in both simulation and physical domain based on the state-of-the-art DNN model as well as the commercial systems. The experimental results suggest that our attack is real and can be launched physically with high attack success rate. 
Our findings real the security issues of DNN in the community of face analysis and we hope our attack brings additional attention on such issues.   
\bibliographystyle{ieeetr} 
\bibliography{ref}

\newpage
\appendix

\subsection{Details of VGGFace Model Optimization}
To fine-tune the VGGFace model \cite{parkhi2015deep}, we set the batch size as $128$, the learning rate with $0.0001$ with weight decay as $0.0005$. We use the Adam algorithm for optimization, where the number of training epochs is set as $100$. The classification accuracy is $0.9428$ and $0.9870$ on the clean data of PubFig and Youtube Face dataset, respectively. 


\subsection{Impact of LED Frequency and Duty Cycle on Face Recognition Task}

\begin{figure}[!t]
    \centering
    \subfigure[]{\includegraphics[width=0.8\columnwidth, trim=110 0 600 0, clip]{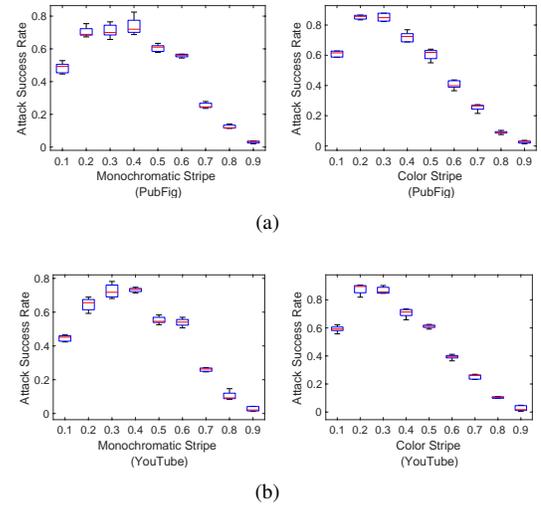}}
    \subfigure[]{\includegraphics[width=0.8\columnwidth, trim=600 0 110 0, clip]{figure/detect_duty.eps}}
    \caption{The impact of duty circle on attack success rate of face recognition task.}
    \label{fig:detect_duty_recognition}
    \vspace{-10pt}
\end{figure}

In this section, we evaluate the LED setting by analyzing the impact of LED frequency and duty cycle on face recognition Task. Noted that for face recognition task, we are able to fine-tune the network based on the trigger samples, which is different from the setting of face verification task. Following the experimental analysis in Section VI, we fix the frequency or duty cycle by setting other parameters obtained through evolutionary computing. We first discuss the performance by varying frequency, where the results are shown in Table VII.  Similar to the observation in Section VI, the attack success rates drop by varying the frequency, which is reasonable as LED frequency is one of the important components to model the stripe pattern, which can have huge impact on the final performance. Nevertheless, the attack success rates are still satisfactory with high face detection success rates and victim success rates. 
We then discuss the impact of duty cycle, where the results are shown in Fig.~\ref{fig:detect_duty_recognition}. As we can observe, we can draw a similar conclusion as verification task despite the fact that the DNN model will be fine-tuned with injected poisoning samples, which is that, the attack performance degrades when either duty cycle is too high or too low as it can influence the quality of captured face images.

\begin{table*}[htbp]
\label{Table: LED_parameter}
\centering

\subtable[Evaluation Result of VGGFace on PubFig Dataset]{
\begin{tabular}{cccccc}
\hline
freq (hz)  & non-victim test accuracy & face detection success rate & victim success rate & attack success rate \\ \hline
50   & 0.9403$\pm$0.0021          & 0.7640$\pm$0.0098 & 0.920$\pm$0.0781 & 0.4398$\pm$0.0832       \\
60   & 0.9411$\pm$0.0016          & 0.7800$\pm$0.0041 & 0.925$\pm$0.0602 & 0.5599$\pm$0.0972        \\
110  & 0.9411$\pm$0.0023          & 0.8280$\pm$0.0028 & 0.850$\pm$0.1140 & 0.8070$\pm$0.0428        \\
220  & 0.9415$\pm$0.0018          & 0.9950$\pm$0.0005 & 0.945$\pm$0.0415 & 0.0648$\pm$0.0405       \\
330  & 0.9409$\pm$0.0023          & 0.9767$\pm$0.0012 & 0.920$\pm$0.0812 & 0.5982$\pm$0.0736       \\ \hline
\end{tabular}
}
\subtable[Evaluation Result of VGGFace on YouTube Face Dataset]{
\begin{tabular}{cccccc}
\hline
freq (hz)  & non-victim test accuracy & {face detection success rate} & victim success rate&  attack success rate \\ \hline
50  & 0.9863$\pm$0.0008 & 0.8003$\pm$0.0037 & 0.975$\pm$0.0433 & 0.0896$\pm$0.0439  \\
60  & 0.9851$\pm$0.0005 & 0.7703$\pm$0.0062 & 0.963$\pm$0.0650 & 0.2848$\pm$0.0924       \\
110 & 0.9862$\pm$0.0014 & 0.6580$\pm$0.0088& 0.950$\pm$0.0707 & 0.5704$\pm$0.0766      \\
220 & 0.9856$\pm$0.0003 & 0.9967$\pm$0.0005 & 1.000$\pm$0.0000 & 0.0561$\pm$0.0114          \\
330 & 0.9854$\pm$0.0007 & 0.9437$\pm$0.0049 & 0.963$\pm$0.0415 & 0.5891$\pm$0.0550  \\ \hline
\end{tabular}
}

\caption{The impact of frequency on attack success rate of face recognition task}
\vspace{-15pt}
\end{table*}

\subsection{Impact of Number of poisoning Samples for Face Recognition Task}

\begin{figure}[htbp]
	\centering
	\subfigure[PubFig dataset]{\includegraphics[width=0.4\columnwidth, trim=50 0 380 0,clip]{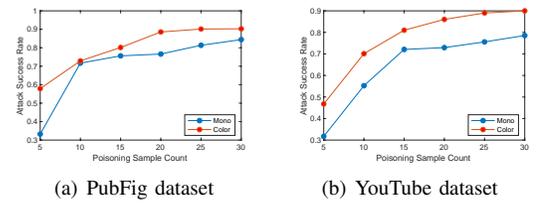}}
	\subfigure[YouTube dataset]{\includegraphics[width=0.4\columnwidth, trim=380 0 50 0,clip]{figure/poison_num.eps}}

	\caption{Attack success rates by considering different number of poisoning samples. }\label{fig:poi}
	\vspace{-10pt}
\end{figure}

In our face recognition experiment in Section VI, we fix the number of poisoning samples as $20$ for both PubFig dataset and Youtube Face dataset. In this section, we are interested in the impact of number of poisoning samples injected into the database.  The results are presented in Fig.~\ref{fig:poi}. We can draw a similar conclusion compared with the observation in \cite{chen2017targeted,yao2019latent} that by injecting more poisoning samples, we can achieve better attack success rates. We find that a relatively stable attack success rates can be achieved when the number of poisoning samples is larger than $20$ for both PubFig and YouTube Face dataset, which corresponds to modulate LED for around $1$ second given the frame rate as $25Hz$. 

\subsection{Physical Implementation based on Household LED}
\label{physicalGray}
\begin{figure}[t]
    \centering
    \subfigure[Simplified circuit]{
    \includegraphics[width=0.6\columnwidth]{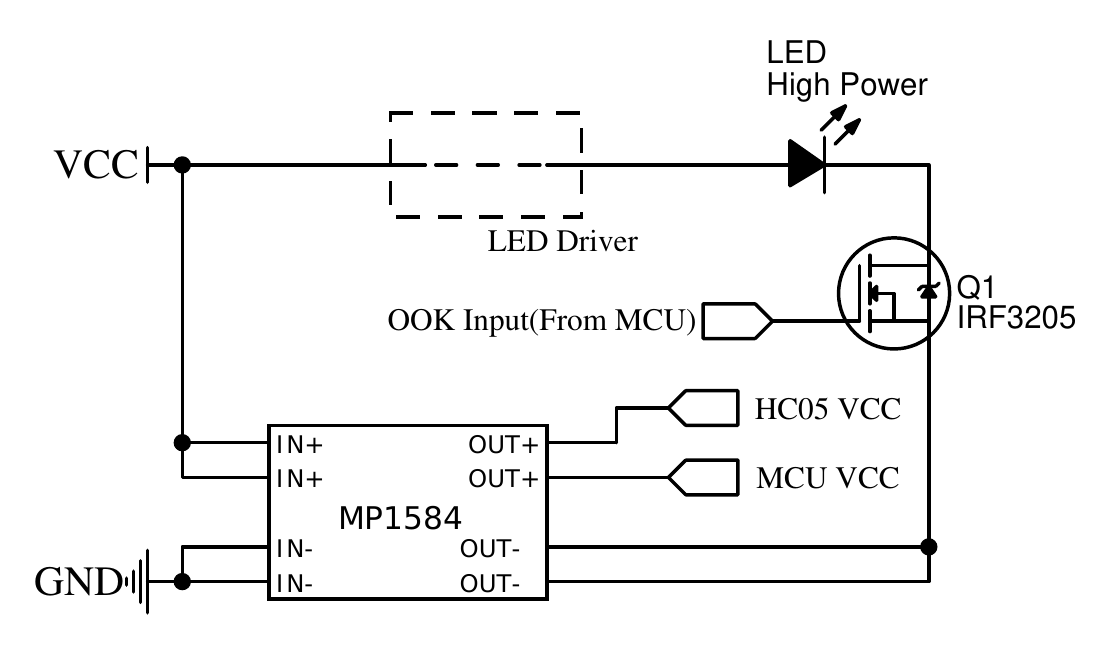}}
    \subfigure[Physical implementation]{
    \includegraphics[width=2cm]{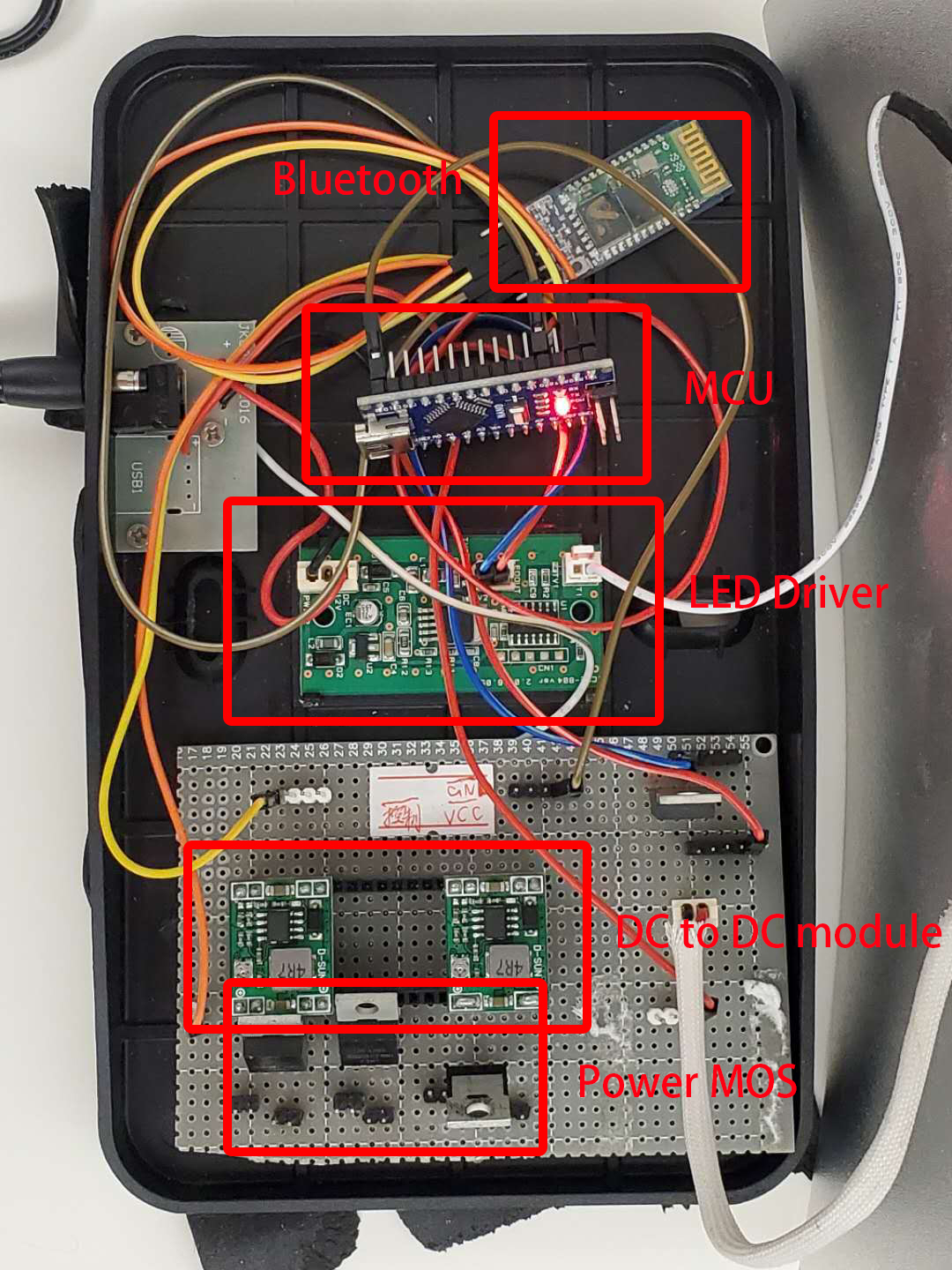}}
    \caption{Implementation based on household LED light.}
    \label{fig:gray_circuit}
    \vspace{-15pt}
\end{figure}

 We use a step-down DC-DC module to provide power for the Micro-Control Unit (MCU) and other peripherals. The input of the step-down switching regulator comes from the LED power supply, thus we can hide the system into a household LED device. More specifically, we adopt MP1584 which is a high frequency step-down switching regulator with an integrated internal high-side high voltage power MOSFET. The input voltage ranges from 4.5V to 28V which accommodates a wide variety of LED supplies. the output voltage can be adjusted from 0.8V to 25V. 

To further modulate LED's ON-OFF keying, we use a Power MOSFET to control the power supply directly connected to the LED. Power MOSFET has a fast switching speed and has a high current capacity. We use IRF3205 to drive the high-power LED. The OOK signal is produced by an Arduino which is a low-cost micro-controller unit. Thus, our implemented system with LED bulb consumes approximately 30W peak power, which is sufficient for an attacker to conduct attack in the office environment 
under normal light condition. 


In addition, a Bluetooth module HC05 is adopted and the I/O pin of Arduino with TTL logic level is directly connected to the gate pin of IRF3205 due to the acceptable low on-resistance, such that an attack can remotely control the LED modulation and configuration without the awareness of victims.


\subsection{Synchronization of Registration and Attack}

\begin{figure}[h]
    \centering
    \includegraphics[width=0.6\columnwidth]{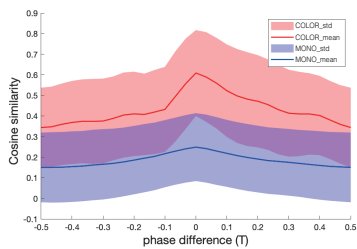}
    \caption{The impact of phase difference on cosine similarity. }
    \label{fig:phase_diff}
\end{figure}

While we can model the intensity, width as well as color information of stripe pattern, the location of the pattern is beyond the control of an attacker, which is due to the synchronization of LED flickering between face registration and attack (see Fig.~\ref{fig:stripe_discuss} for example). Intuitively, if the position of the stripes on the two faces are similar, the face pair should have a higher similarity score. However, in practice, it is highly likely there exists ``phase shift" between the face images for registration and attack purpose. To better understand how synchronization issue influences the final performance, we conduct experiment by varying the phase difference between the registration and attack face images on LFW dataset. We report the mean and standard deviation of cosine similarity of images which belong to two different identities.  The results are shown in Fig. \ref{fig:phase_diff}, where $T$ denotes the period of LED flickering. As we can observe, two faces images are more similar when there is no phase difference between them, and the similarity drops when phase difference becomes larger. As an attacker cannot control the phase difference, we therefore recommend to conduct multiple trials, which is allowed by off-the-shelf face recognition systems.  

\subsection{LED Phase Shift}
\begin{figure}[t]
    \centering
    \includegraphics[width=0.55\columnwidth, trim=0 0 0 0, clip]{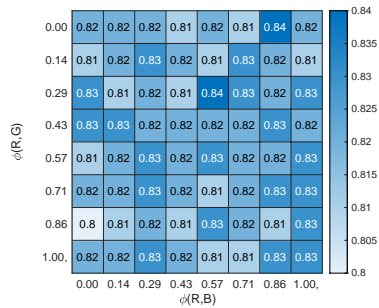}
    \caption{The impact of phase shift on attack success rate.}
    \label{fig:phase_heatmap}
\end{figure}

We are also interested in the phase shift among different LED sources. To this end, we conduct experiments by varying the phase shift of $\phi(R,G)$ and $\phi(R,B)$ in terms of channel R.  As we can observe from Fig.~\ref{fig:phase_heatmap}, the performance vary by choosing different phase shift, which shows that the DNN model for face analysis can be impacted by color information. Nevertheless, we find that our proposed attack strategy can achieve significantly better performance by randomly selecting color combination in general, which further indicates the effectiveness of our proposed attack framework.

\subsection{Impact of the stripe Direction}
\begin{figure}[!t]
    \centering
    \subfigure[Monochromatic stripe patterns]{\includegraphics[width=4cm,trim=40 0 300 0, clip]{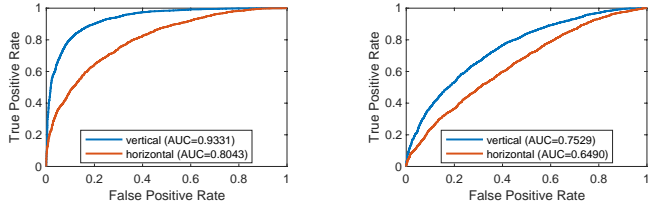}}
    \qquad
    \subfigure[Color stripe patterns]{\includegraphics[width=4cm,trim=300 0 40 0, clip]{figure/direction.eps}}
    \caption{The impact of stripe direction on ROC curve.} 
    \label{fig:direction}
\end{figure}

While in the previous study we assume that stripe patterns are generated in a column-wise manner, it may not always be true due to the orientation of camera as well as the design of CMOS sensor. To this end, we discuss the impact of stripe direction by further considering stripe patterns generated in a row-wise manner, where the parameters are optimized based on CMA-ES. The results are shown in Fig. \ref{fig:direction}. Surprisingly, we find that better attack performance can be achieved by injecting horizontal stripe pattern. We conjecture the reason 
may due to the structural symmetry of human face, such that the DNN model preserve more horizontal information.

\subsection{Quality of Face Images}

We finally discuss the quality of captured face images with stripe patterns. Noted that different APIs may adopt different metrics, such that the evaluation can be different. Besides directly measuring the quality based on whether the face region can be detected or not, we also consider to use the quality metrics adopted by Baidu API, which mainly evaluates the quality from three different perspectives, namely occlusion, blur and illumination. We use the LFW dataset by modeling the stripe patterns based on the parameters through evolutionary computing. We list the averaging score obtained through Baidu API as well as its recommended threshold in Table \ref{tab:quality}. As we can see, our proposed backdoor trigger will not impact the quality, which guarantees reliable face detection rate.

\begin{table}[!t]
\caption{Face image quality based on Baidu API. }
\centering
\begin{tabular}{clcc}

\toprule
\multicolumn{2}{c}{Evaluation index} & score & recommended threshold \\ \hline
\multirow{7}{*}{occulusion} & left\_eye & 0.06 & \textless{}0.6 \\
 & right\_eye & 0.15 & \textless{}0.6 \\
 & nose & 0.19 & \textless{}0.7 \\
 & mouth & 0.16 & \textless{}0.7 \\
 & left cheek & 0.11 & \textless{}0.8 \\
 & right\_cheek & 0.26 & \textless{}0.8 \\
 & chin\_contour & 0.21 & \textless{}0.6 \\
 \hline
\multicolumn{2}{c}{blur} & 0.08 & \textless{}0.7 \\
\hline
\multicolumn{2}{c}{illumination} & 124 & \textgreater{}40 \\ 
\bottomrule
\end{tabular}
\label{tab:quality}
\end{table}




\end{document}